\documentclass[superscriptaddress,onecolumn,times,notitlepage]{revtex4-1}
\usepackage{amsfonts}
\usepackage{amssymb,amsmath}
\usepackage{graphicx}
\usepackage{subfigure}
\usepackage{color}
\usepackage[normalem]{ulem}
\usepackage{natbib}
\usepackage{bbold}
\usepackage{placeins}
\usepackage{soul,xcolor}
\usepackage{epstopdf}
\usepackage{tabu}
\usepackage{array}
\usepackage{upgreek}
\usepackage{braket}
\DeclareMathAlphabet{\mathpzc}{OT1}{pzc}{m}{it}
\usepackage{xcolor}

\usepackage{multirow}
\usepackage[utf8]{inputenc}
\usepackage[colorlinks = truelinkcolor = blue, urlcolor  = blue, citecolor = blue, anchorcolor = blue]{hyperref}
\hypersetup{
	colorlinks   = true, 
	urlcolor     = blue, 
	linkcolor    = blue, 
	citecolor   = blue 
}

\begin{document}
\title{Quantum phase properties of photon added and subtracted displaced
Fock states}

	\author{Priya Malpani}
	\affiliation{Indian Institute of Technology Jodhpur, Jodhpur 342037, India}
	\author{Kishore Thapliyal}
	\affiliation{RCPTM, Joint Laboratory of Optics of Palacky University and Institute of Physics of Academy of Science of the Czech Republic, Faculty of Science, Palacky University, 17. listopadu 12, 771 46 Olomouc, Czech Republic}
	\affiliation{Jaypee Institute of Information Technology, A-10, Sector-62, Noida UP-201309, India}
    \author{Nasir Alam}
    \affiliation{Jaypee Institute of Information Technology, A-10, Sector-62, Noida UP-201309, India}
    \author{Anirban Pathak}
    \affiliation{Jaypee Institute of Information Technology, A-10, Sector-62, Noida UP-201309, India}
	\author{V. Narayanan}
	\affiliation{Indian Institute of Technology Jodhpur, Jodhpur 342037, India}
	\author{Subhashish Banerjee}
	\affiliation{Indian Institute of Technology Jodhpur, Jodhpur 342037, India}

%
\begin{abstract}
Quantum phase properties of photon added and subtracted displaced
Fock states (and a set of quantum states which can be obtained as
the limiting cases of these states) are investigated from a number
of perspectives, and it is shown that the quantum phase properties
are dependent on the quantum state engineering operations performed.
Specifically, the analytic expressions for quantum phase distributions
and angular $Q$ distribution as well as measures of quantum phase
fluctuation and phase dispersion are obtained. The uniform phase distribution
of the initial Fock states is observed to be transformed by the unitary
operation (i.e., displacement operator) into non-Gaussian shape, except
for the initial vacuum state. It is observed that the phase distribution
is symmetric with respect to the phase of the displacement parameter
and becomes progressively narrower as its amplitude increases. The
non-unitary (photon addition/subtraction) operations make it even narrower in contrast to the Fock parameter, which leads to
broadness. The photon subtraction is observed to be a more powerful
quantum state engineering tool in comparison to the photon addition.
Further, one of the quantum phase fluctuation parameters is found
to reveal the existence of antibunching in both the engineered quantum
states under consideration. Finally, the relevance of the engineered
quantum states in the quantum phase estimation is also discussed,
and photon added displaced Fock state is shown to be preferable for
the task. 
\end{abstract}
\maketitle
\section{Introduction}

Generation of the desired nonclassical states (i.e., quantum states
without any classical counterpart), characterized by negative Glauber-Sudarshan
$P$ function \cite{glauber1963coherent,sudarshan1963equivalence},
from a given initial state can be performed with the help of both
unitary dynamics and non-unitary operations. The unitary time evolution
under the control of a Hamiltonian may result in the nonclassical
states \cite{thapliyal2014higher,thapliyal2015quasiprobability,thapliyal2014nonclassical,giri2017nonclassicality,bachor2004guide},
but experimentally this is often limited to a small set of implementable
Hamiltonians. Another approach, which includes non-unitary measurement
or photon addition/subtraction operations, is known to be useful in
some cases for generation of such states. A general recipe including
both unitary and non-unitary operations, which enables us to generate
a large set of nonclassical states in experiments, is labeled as quantum
state engineering \cite{vogel1993quantum,sperling2014quantum,miranowicz2004dissipation,marchiolli2004engineering,escher2004controlled,yukawa2013generating,marek2018loop}. 

Particularly interesting examples of such engineered nonclassical
states are Fock state, photon added/subtracted coherent state \cite{agarwal1991nonclassical},
displaced Fock state (DFS) \cite{satyanarayana1985generalized,wunsche1991displaced,ziesel2013experimental,zavatta2004quantum},
photon added DFS (PADFS), and photon subtracted DFS (PSDFS) \cite{malpani2019lower}.
Here, we address the relevance of such engineered quantum states in
implementation of different tasks exploiting their phase properties. 
Our specific interest is to study the phase properties of PADFS and
PSDFS, also known as photon added and subtracted generalized coherent
states \cite{malpani2019lower}, as a large set of engineered quantum
states (along with the conventional coherent state) can be obtained
in the limiting cases of PADFS and PSDFS. In the recent past, the
nonclassical properties of this set of engineered quantum states,
many of which have been experimentally generated \cite{lvovsky2001quantum,lvovsky2002synthesis,zavatta2004quantum,zavatta2005single,zavatta2008subtracting},
were  focus of various studies (see
\cite{malpani2019lower} and references therein). However, in the
present work, we do not want to focus on the specific nonclassical
properties of these states. Rather, we wish to
investigate the phase properties of these states for the reasons explained
below.

The impossibility of writing a Hermitian operator for quantum phase is
a longstanding problem (see \cite{perinova1998phase,carruthers1968phase,lynch1987phase}
for review). Early efforts of Dirac \cite{dirac1927quantum} to introduce
a Hermitian quantum phase operator were not successful, but led to
many interesting proposals \cite{susskind1964quantum,pegg1989phase,barnett1986phase}.
Specifically, Susskind-Glogower \cite{susskind1964quantum}, Pegg-Barnett
\cite{pegg1988unitary,pegg1989phase,barnett1990quantum}, and Barnett-Pegg \cite{barnett1986phase} formalisms
played very important role in the studies of phase properties and
the phase fluctuation \cite{imry1971relevance}. Thereafter, phase
properties of various quantum states have been reported using these
formalisms \cite{sanders1986bc,gerry1987phase,yao1987phase,carruthers1968phase,vaccaro1989phase,pathak2000phase,alam2016quantum,alam2017quantum,verma2009reduction}.
Other approaches have also been used for the study of the phase properties.
For example, quantum phase distribution is defined using phase states
\cite{agarwal1992classical}, while Wigner \cite{garraway1992quantum}
and $Q$ \cite{leonhardt1993phase,leonhardt1995canonical} phase distributions
are obtained by integrating over radial parameter of the corresponding
quasidistribution function. In experiments, the phase measurement
is obtained by averaging the field amplitudes of the $Q$ function
\cite{noh1991measurement,noh1992operational}; Pegg-Barnett and Wigner
phase distributions are also reported with the help of reconstructed
density matrix \cite{smithey1993complete}. Further, quantum phase
distribution under the effect of the environment was also studied in the
past leading to phase diffusion \cite{banerjee2007phase,banerjee2007phaseQND,abdel2010anabiosis,banerjee2018open}.
A measure of phase fluctuation named phase dispersion using quantum
phase distribution has also been proposed in the past \cite{perinova1998phase,banerjee2007phase}.
Recently, quantum phase fluctuation \cite{zheng1992fluctuation} and
Pancharatnam phase \cite{mendas1993pancharatnam} have been studied
for DFS. The quantum phase fluctuation in parametric down-conversion \cite{gantsog1991quantum}
and its revival \cite{gantsog1992collapses} are also reported. Experiments
on phase super-resolution without using entanglement \cite{resch2007time}
and role of photon subtraction in concentration of phase information
\cite{usuga2010noise} are also performed. Optimal phase estimation \cite{sanders1995optimal}
using different quantum states \cite{higgins2007entanglement} (including
NOON and other entangled states and unentangled single-photon states)
has long been the focus of quantum metrology \cite{giovannetti2006quantum,giovannetti2011advances}.
More recently, some efforts have been made to describe quasidistribution
of phase as filtering in the interval of phase difference \cite{perina2019quasidistribution}.
In brief, quantum phase properties are of intense interest of the
community since long (see \cite{pathak2002quantum,perinova1998phase}
and references therein), and the interest in
it has been further intensified in the recent past as many new applications
of quantum phase distribution and quantum phase fluctuation have been
realized.

To stress on the recently reported applications of quantum phase distribution
and quantum phase fluctuation, we note that these have applications
in quantum random number generation \cite{xu2012ultrafast,raffaelli2018soi},
cryptanalysis of squeezed state based continuous variable quantum
cryptography \cite{horak2004role}, generation of solitons in a Bose-Einstein
condensate \cite{denschlag2000generating}, storage and retrieval
of information from Rydberg atom \cite{ahn2000information}, in phase
encoding quantum cryptography \cite{gisin2002quantum}, phase imaging
of cells and tissues for biomedical application \cite{park2018quantitative};
as well as have importance in determining the value of transition
temperature for superconductors \cite{emery1995importance}. Keeping
these applications and the general nature of engineered quantum states
PADFS and PSDFS in mind, we study phase distribution, $Q$ phase,
phase fluctuation measures, phase dispersion, and quantum phase estimation
using the concerned states and the states obtained in the limiting
cases. 

The rest of the paper is organized as follows. In Section \ref{sec:Quantum-phase-parameters},
we briefly introduce various parameters (e.g., quantum phase distribution,
angular $Q$ function, quantum phase fluctuation measures, and uncertainty
in quantum phase estimation) which are used in this paper to study
the quantum phase properties. In Section \ref{sec:Quantum-states-of},
we describe the PADFS, PSDFS, and a set of quantum states that can
be obtained in the limiting cases. In Section \ref{sec:phase-witnesses},
we investigate the phase properties of PADFS and PSDFS from a number
of perspectives. Finally, the paper is concluded in Section \ref{sec:Conclusions}.

\section{Quantum phase distribution and other phase properties \label{sec:Quantum-phase-parameters}}

Quantum phase operator $\hat{\phi}$ was introduced by Dirac based
on his assumption that the annihilation operator $\hat{a}$ can be
factored out into a Hermitian function $f(\hat{N})$ of the number
operator $\hat{N}=\hat{a}^{\dagger}\hat{a}$ and a unitary operator
$\hat{U}$ \cite{dirac1927quantum} as

\begin{equation}
\hat{a}=\hat{U}\,f\left(\hat{N}\right),\label{eq:Dirac_pahse}
\end{equation}
where
\begin{equation}
\hat{U}=e^{\iota\hat{\phi}}.\label{eq:phase-operator}
\end{equation}
However, there was a problem with the Dirac formalism of phase operator
as it failed to provide a meaning to the corresponding uncertainty
relation. Specifically, in the Dirac formalism, the creation ($\hat{a}^{\dagger}$)
and annihilation ($\hat{a}$) operators satisfy the bosonic commutation
relation, $\left[\hat{a},\,\hat{a}^{\dagger}\right]=1$, iff $\left[\hat{N},\,\hat{\phi}\right]=\iota$,
which leads to the number phase uncertainty relation $\Delta N\,\Delta\phi\geq1$.
Therefore, in order to satisfy the bosonic commutation relation under
Dirac formalism, the phase uncertainty should be greater than 2$\pi$
for $\Delta N$ < $\frac{1}{2\pi}$ which lacks a physical description.
Subsequently, Louisell \cite{louisell1963amplitude} proposed some
periodic phase based method, which was followed by Susskind and Glogower
formalism based on Sine and Cosine operators \cite{susskind1964quantum}.
An important contribution to this problem is the Barnett-Pegg
formalism \cite{barnett1986phase} which is used in this work. In what
follows, we will also briefly introduce notions, such as quantum phase
distribution, angular $Q$ phase function, phase fluctuation parameters,
phase dispersion, quantum phase estimation to study the phase properties
of the quantum states of our interest. 

\subsection{Quantum phase distribution}

A distribution function allows us to calculate expectation values
of an operator analogous to that from the corresponding density matrix.
Phase distribution function for a given density operator \cite{banerjee2007phase,agarwal1992classical}
can be defined as

\begin{equation}
P_{\theta}=\frac{1}{2\pi}\langle\theta|\varrho|\theta\rangle,\label{eq:Phase-Distridution}
\end{equation}
where the phase state $|\theta\rangle$, complementary to the number state
$|n\rangle$, is defined \cite{agarwal1992classical} as

\begin{equation}
|\theta\rangle=\sum_{n=0}^{\infty}e^{\iota n\theta}|n\rangle.\label{eq:phase}
\end{equation}
From the definition of the phase distribution (\ref{eq:Phase-Distridution}),
it can be observed that for a Fock state, $P_{\theta}=\frac{1}{2\pi}$,
implying it has a uniform distribution of phase. Interestingly, the
states of our interest, PADFS and PSDFS, are obtained by displacing
the Fock state followed by photon addition/subtraction. Therefore,
we will study here what is the effect of application of displacement
operator on a uniformly phase distributed (Fock) state and how subsequent
photon addition/subtraction further alters the phase distribution.
Using phase distribution function, the information regarding uncertainty
in phase and phase fluctuation can also be obtained. 

\subsection{Angular $Q$ phase function}

Analogous to the phase distribution $P_{\theta}$, phase distributions
are also defined as radius integrated quasidistribution functions
which are used as the witnesses for quantumness \cite{thapliyal2015quasiprobability}.
One such phase distribution function based on the angular part of
the $Q$ function is studied in \cite{leonhardt1993phase,leonhardt1995canonical}.
Specifically, the angular $Q$ function is defined as

\begin{equation}
Q_{\theta_{1}}=\intop_{0}^{\infty}Q\left(\beta,\beta^{\star}\right)\left|\beta\right|d\left|\beta\right|,\label{eq:ang-Qf}
\end{equation}
where the $Q$ function \cite{husimi1940some} is defined as the projection
of the state of interest on the coherent state basis, i.e.,

\begin{equation}
Q=\dfrac{1}{\pi}\langle\beta|\rho|\beta\rangle\label{eq:Qf}
\end{equation}
with coherent state parameter $\beta=\left|\beta\right|\exp\left[\iota\theta_{1}\right]$.
The relevance of the $Q$ function as witness of nonclassicality \cite{thapliyal2015quasiprobability}
and in state tomography \cite{thapliyal2016tomograms} is well studied.
On top of that, non-Gaussianity of the PADFS and PSDFS using $Q$
function was recently reported by us \cite{malpani2019lower}.

\subsection{Quantum phase fluctuation}

In attempts to get rid of the limitations of the Hermitian phase operator
of Dirac \cite{dirac1927quantum}, Louisell \cite{louisell1963amplitude}
first mentioned that bare  phase operator should be replaced by periodic functions.
As a consequence, sine $(\mathcal{\hat{S}})$ and cosine $(\hat{\mathcal{C}})$
operators appeared, explicit forms of these operators were given by
Susskind and Glogower \cite{susskind1964quantum}, and further modified
by Barnett and Pegg \cite{barnett1986phase} as
\begin{equation}
\mathcal{\hat{S}}=\frac{\hat{a}-\hat{a}^{\dagger}}{2\iota\left(\bar{N}+\frac{1}{2}\right)^{\frac{1}{2}}}\label{eq:fluctuation1}
\end{equation}
and 
\begin{equation}
\hat{\mathcal{C}}=\frac{\hat{a}+\hat{a}^{\dagger}}{2\left(\bar{N}+\frac{1}{2}\right)^{\frac{1}{2}}}.\label{eq:fluctuation2}
\end{equation}
Here, $\bar{N}$ is the average number of photons in the measured
field, and here we refrain our discussion to Barnett and Pegg sine and cosine operators \cite{barnett1986phase}. Carruthers and Nieto \cite{carruthers1968phase} have introduced
three quantum phase fluctuation parameters in terms of sine and cosine
operators 
\begin{equation}
U=\left(\Delta N\right)^{2}\left[\left(\Delta\mathcal{S}\right)^{2}+\left(\Delta\mathcal{C}\right)^{2}\right]/\left[\langle\mathcal{\hat{S}}\rangle^{2}+\langle\hat{\mathcal{C}}\rangle^{2}\right],\label{eq:fluctuation3}
\end{equation}
\begin{equation}
S=\left(\Delta N\right)^{2}\left(\Delta\mathcal{S}\right)^{2},\label{eq:fluctuation4}
\end{equation}
and
\begin{equation}
Q=S/\langle\hat{\mathcal{C}}\rangle^{2}.\label{eq:fluctuation5}
\end{equation}
Note that Carruthers and Nieto \cite{carruthers1968phase} had introduced these parameters in terms of Susskind and Glogower operators \cite{susskind1964quantum}; here we use them in Barnett-Pegg formalism to remain consistent with \cite{gupta2007reduction}, where $U$ parameter is shown relevant as a witness of nonclassicality \cite{gupta2007reduction}. Specifically, $U$ is 0.5 for coherent state, and reduction
of $U$ parameter below the value for coherent state can be interpreted
as the presence of nonclassical behavior \cite{gupta2007reduction}.
In what follows, we will study quantum phase fluctuations for PADFS
and PSDFS by computing analytic expressions of $U,\,S$ and $Q$ parameters
in Barnett-Pegg formalism, with a specific focus
on the possibility of witnessing nonclassical properties of these
states via the reduction of $U$ parameter below the coherent state
limit.

\subsection{Phase dispersion}

A known application of phase distribution function (\ref{eq:Phase-Distridution})
is that it can be used to quantify the quantum phase fluctuation.
Although the variance is also used occasionally as a measure of phase
fluctuation,  it has a drawback that
it depends on the origin of phase integration \cite{banerjee2007phase}.
A measure of phase fluctuation, free from this problem, is phase dispersion
\cite{perinova1998phase} defined as

\begin{equation}
D=1-\left|\intop_{-\pi}^{\pi}d\theta\exp\left[-\iota\theta\right]P_{\theta}\right|^{2}.\label{eq:Dispersion}
\end{equation}
Here, it is worth stressing that both Carruthers-Nieto parameters
and phase dispersion $D$ correspond to phase fluctuation, Our primary
focus is to study phase fluctuation and further to check the correlation
between these measures of phase fluctuation. Thus, it would be interesting
to study phase fluctuation from the two perspectives.

\subsection{Quantum phase estimation}

Quantum phase estimation is performed by sending the input state through
a Mach-Zehnder interferometer and applying the phase to be determined
($\phi$) on one of the arms of the interferometer. To study the phase
estimation using Mach-Zehnder interferometer, angular momentum operators
\cite{sanders1995optimal,demkowicz2015quantum}, defined as

\begin{equation}
\hat{J_{x}}=\frac{1}{2}\left(\hat{a}^{\dagger}\hat{b}+\hat{b}^{\dagger}\hat{a}\right),\label{eq:ang-mom1}
\end{equation}

\begin{equation}
\hat{J_{y}}=\frac{\iota}{2}\left(\hat{b}^{\dagger}\hat{a}-\hat{a}^{\dagger}\hat{b}\right),\label{eq:ang-mom2}
\end{equation}
and

\begin{equation}
\hat{J_{z}}=\frac{1}{2}\left(\hat{a}^{\dagger}\hat{a}-\hat{b}^{\dagger}\hat{b}\right),\label{eq:ang-mom3}
\end{equation}
are used. Here, $\hat{a}$ and $\hat{b}$ are the annihilation operators
for the modes corresponding to two input ports of the Mach-Zehnder
interferometer. The average value of $\hat{J_{z}}$ operator in the
output of the Mach-Zehnder interferometer, which is one-half of the
difference of photon numbers in the two output ports (\ref{eq:ang-mom3}),
can be written as 
\begin{equation}
\langle\hat{J_{z}}\rangle=\cos\text{\ensuremath{\phi}}\langle\hat{J_{z}}\rangle_{in}-\sin\text{\ensuremath{\phi}}\langle\hat{J_{x}}\rangle_{in}.
\end{equation}
Therefore, variance in the measured value of operator $\hat{J_{z}}$
can be computed as
\begin{equation}
\left(\Delta J_{z}\right)^{2}=\cos^{2}\phi\left(\Delta{J_{z}}\right)_{in}^{2}+\sin^{2}\phi\left(\Delta{J_{x}}\right)_{in}^{2}-2\sin\text{\ensuremath{\phi\,\cos\phi\,}cov}\left(\hat{J_{x}},\hat{J_{z}}\right)_{in},
\end{equation}
where covariance of the two observables is defined as
\begin{equation}
\text{{\rm cov}}\left(\hat{J_{x}},\hat{J_{z}}\right)=\frac{1}{2}\langle\hat{J_{x}}\hat{J_{z}}+\hat{J_{z}}\hat{J_{x}}\rangle-\langle\hat{J_{x}}\rangle\langle\hat{J_{z}}\rangle.
\end{equation}
This allows us to quantify precision in phase estimation \cite{demkowicz2015quantum}
as

\begin{equation}
\Delta\phi=\frac{\Delta{J_{z}}}{\left|\frac{d\langle\hat{J_{z}}\rangle}{d\phi}\right|}.\label{eq:PE}
\end{equation}
In what follows, we have also computed and analyzed $\Delta\phi$
for the quantum states of our interest.

In this section, we have introduced a number of interesting parameters/measures
that can be used to study the phase properties of a quantum state.
In the next section, we will briefly introduce the quantum states
of our interest and in the subsequent section, we aim to study their
phase properties. 

\section{Quantum states of our interest \label{sec:Quantum-states-of}}

We have already mentioned that our focus would be on PADFS and PSDFS.
In the Fock basis, a DFS may be described as \cite{satyanarayana1985generalized}

\begin{equation}
\begin{array}{lcl}
|\Phi(n,\alpha)\rangle=\hat{D}(\alpha)|n\rangle & = & \frac{N}{\sqrt{n!}}\sum\limits _{p=0}^{n}{n \choose p}(-\alpha^{\star})^{(n-p)}\exp\left[-\frac{\mid\alpha\mid^{2}}{2}\right]\sum\limits _{m=0}^{\infty}\frac{\alpha^{m}}{m!}\sqrt{(m+p)!}|m+p\rangle,\end{array}\label{eq:GCS}
\end{equation}
where the displacement operator $\hat{D}(\alpha)=\exp\left(\alpha\hat{a}^{\dagger}-\alpha^{\star}\hat{a}\right)$
is defined in terms of displacement parameter $\alpha=\left|\alpha\right|\exp\left(\iota\theta_{2}\right)$,
and $|n\rangle$ is the initial Fock state chosen to be displaced
with $n$ being the Fock parameter. Using this state,  $u$-PADFS can
be defined as \cite{malpani2019lower}

\begin{equation}
|\psi_{+}(u,n,\alpha)\rangle=N_{+}\hat{a}^{\dagger u}|\Phi(n,\alpha)\rangle=\frac{N_{+}}{\sqrt{n!}}\sum_{p=0}^{n}{n \choose p}(-\alpha^{\star})^{(n-p)}\exp\left(-\frac{\mid\alpha\mid^{2}}{2}\right)\sum_{m=0}^{\infty}\frac{\alpha^{m}}{m!}\sqrt{(m+p+u)!}|m+p+u\rangle.\label{eq:PADFS}
\end{equation}
Similarly, we can define  $v$-PSDFS as \cite{malpani2019lower}

\begin{equation}
|\psi_{-}(v,n,\alpha)\rangle=N_{-}\hat{a}^{v}|\Phi(n,\alpha)\rangle=\frac{N_{-}}{\sqrt{n!}}\sum_{p=0}^{n}{n \choose p}(-\alpha^{\star})^{(n-p)}\exp\left(-\frac{\mid\alpha\mid^{2}}{2}\right)\sum_{m=0}^{\infty}\frac{\alpha^{m}}{m!}\dfrac{(m+p)!}{\sqrt{(m+p-v)!}}|m+p-v\rangle.\label{eq:PSDFS}
\end{equation}
Here, $m$ and $p$ are the real integers. Further, normalization
constants for PADFS and PSDFS can be obtained as
\[
\begin{array}{lcl}
N_{+} & = & \left[\frac{1}{n!}\sum\limits _{p,p'=0}^{n}{n \choose p}{n \choose p'}(-\alpha^{\star})^{(n-p)}(-\alpha)^{(n-p')}\exp\left[-\mid\alpha\mid^{2}\right]\right.\left.\sum\limits _{m=0}^{\infty}\frac{\alpha^{m}(\alpha^{\star})^{m+p-p'}(m+p+u)!}{m!(m+p-p')!}\right]^{-0.5}\end{array}
\]
and 
\[
\begin{array}{lcl}
N_{-} & = & \left[\frac{1}{n!}\sum\limits _{p,p'=0}^{n}{n \choose p}{n \choose p'}\left(\alpha^{\star}\right)^{(n-p)}\left(-\alpha\right){}^{(n-p')}\exp\left[-\mid\alpha\mid^{2}\right]\right.\left.\sum\limits _{m=0}^{\infty}\frac{\alpha^{m}\left(\alpha^{\star}\right){}^{m+p-p'}(m+p)!}{m!(m+p-p')!(m+p-v)!}\right]^{-0.5},\end{array}
\]
respectively. Here, the subscripts $+$ and $-$ correspond to the
photon addition and subtraction, respectively. A schematic scheme
for generation of PADFS/PSDFS from DFS has been proposed recently
by us (see Fig. 1 in Ref. \cite{malpani2019lower}). Due to the general
form of PADFS and PSDFS, a large number of states can be obtained
in the limiting cases. Some of the important limiting cases of PADFS
and PSDFS in the present notation are summarized in Table \ref{tab:state}.
This table clearly establishes that the applicability of the results
obtained in the present study is not restricted to PADFS and PSDFS;
rather an investigation of the phase properties of PADFS and PSDFS
would also reveal phase properties of many other quantum states of
particular interest.

\begin{table}
\begin{centering}
\begin{tabular}{c>{\centering}p{3cm}c>{\centering}p{3cm}}
\hline 
Reduction of state & Name of the state & Reduction of state & Name of the state\tabularnewline
\hline 
\hline 
$|\psi_{+}(u,n,\alpha)\rangle$ & $u$-PADFS & $|\psi_{-}(v,n,\alpha)\rangle$ & $v$-PSDFS\tabularnewline
|$\psi_{+}(0,n,\alpha)\rangle$ & DFS & |$\psi_{-}(0,n,\alpha)\rangle$ & DFS\tabularnewline
$|\psi_{+}(0,0,\alpha)\rangle$ & Coherent state & $|\psi_{-}(0,0,\alpha)\rangle$ & Coherent state\tabularnewline
|$\psi_{+}(0,n,0)\rangle$ & Fock state & |$\psi_{-}(0,n,0)\rangle$ & Fock state\tabularnewline
|$\psi_{+}(u,0,\alpha)\rangle$ & $u$-Photon added coherent state  & |$\psi_{-}(v,0,\alpha)\rangle$ & $v$-Photon subtracted coherent state \tabularnewline
\hline 
\end{tabular}
\par\end{centering}
\caption{\label{tab:state}Various states that can be obtained as the limiting
cases of the PADFS and PSDFS.}
\end{table}

\section{Phase properties of PADFS and PSDFS \label{sec:phase-witnesses}}

The description of the states of our interest given in the previous
section can be used to study different phase properties and quantify
phase fluctuation in the set of quantum states listed in Table \ref{tab:state}.
Specifically, with the help of the quantum states defined in Eqs.
(\ref{eq:PADFS})-(\ref{eq:PSDFS}), we have obtained the analytic
expressions of phase distribution and other phase parameters defined
in Section \ref{sec:Quantum-phase-parameters}.

\subsection{Phase distribution function}

\begin{figure}
\centering{}

\subfigure[]{\includegraphics[scale=0.5]{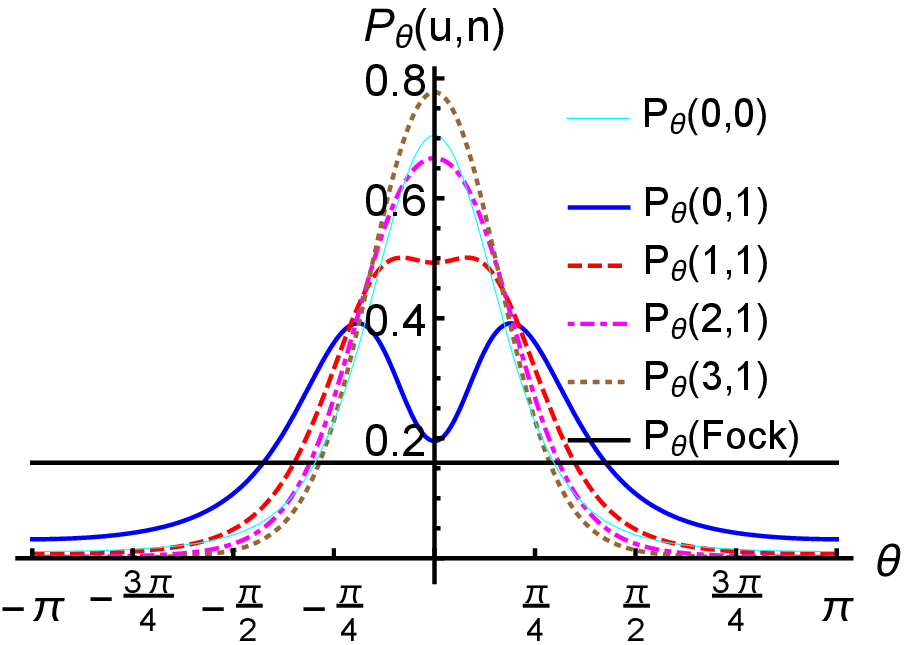}}
\quad{}\quad{}\subfigure[]{ \includegraphics[scale=0.5]{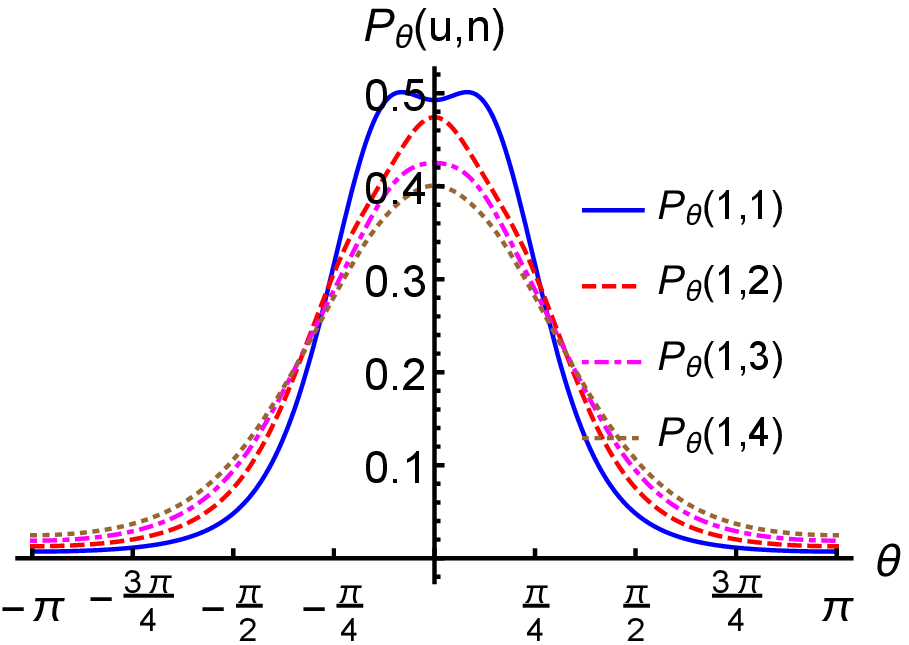}}\\
 \subfigure[]{\includegraphics[scale=0.4]{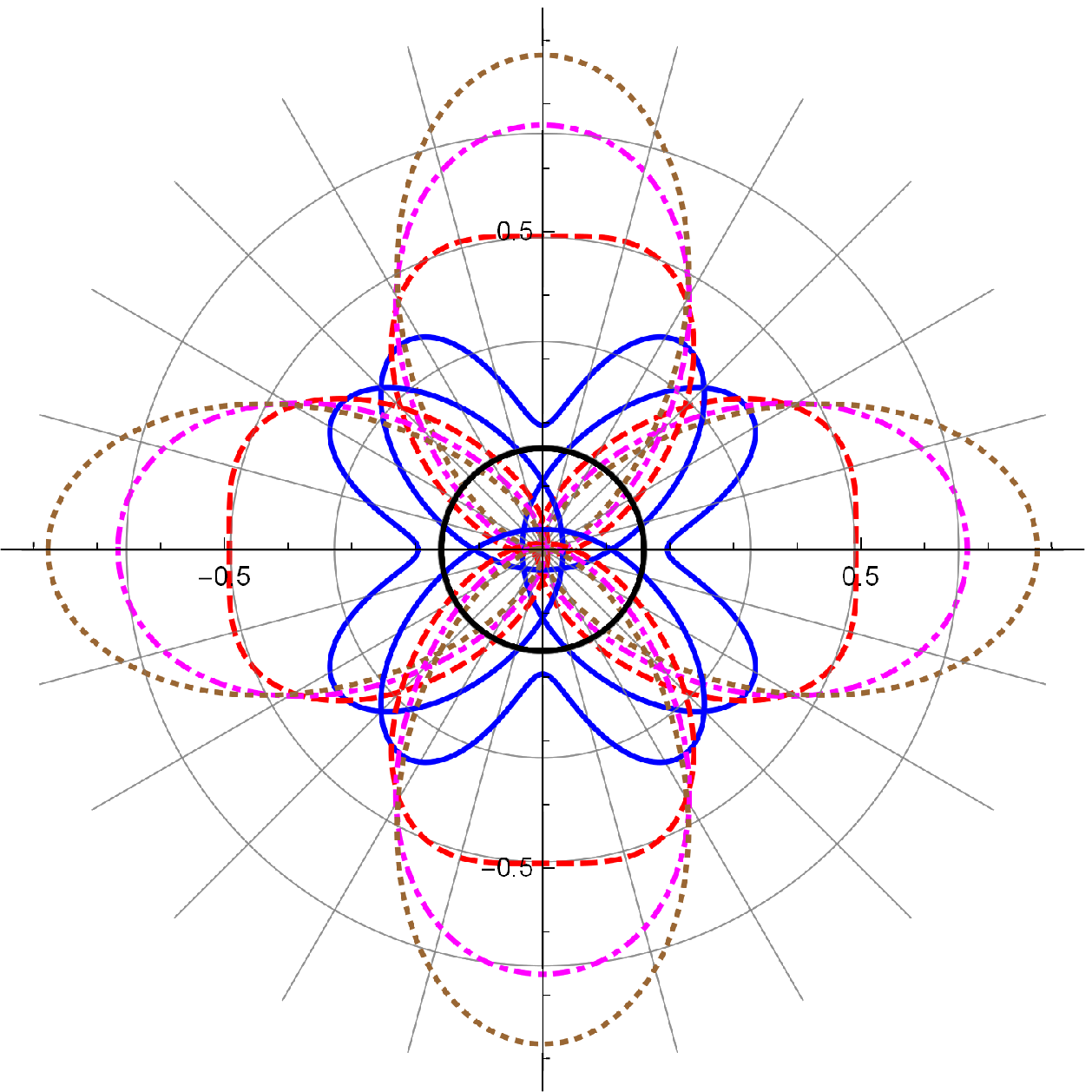}}\quad{}\quad{}\subfigure[]{
\includegraphics[scale=0.4]{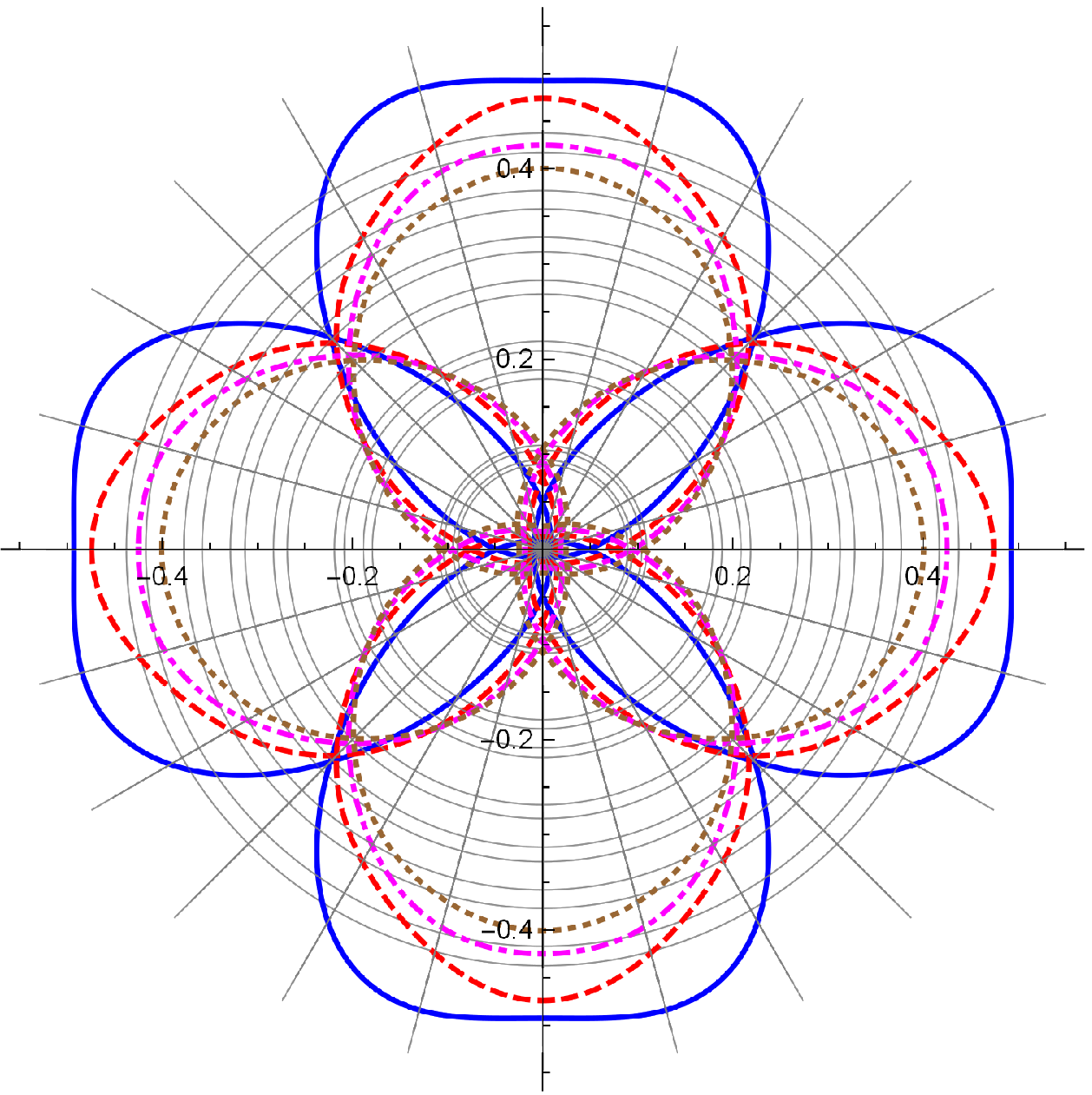}}
\caption{\label{fig:Phase-Distribution-Function} (Color online) Variation
of phase distribution function with phase parameter for PADFS with
displacement parameter $\left|\alpha\right|=1$ for different values
of photon addition ((a) and (c)) and Fock parameters ((b) and (d)).
The phase distribution is shown using both two-dimensional ((a) and
(b) with $\theta_{2}=0$) and polar ((c) and (d)) plots. In (c) and
(d), $\theta_{2}=\frac{n\pi}{2}$ with integer $n\in\left[0,3\right]$,
and the legends are same as in (a) and (b), respectively. }
\end{figure}

\begin{figure}
\centering{}

\subfigure[]{\includegraphics[scale=0.5]{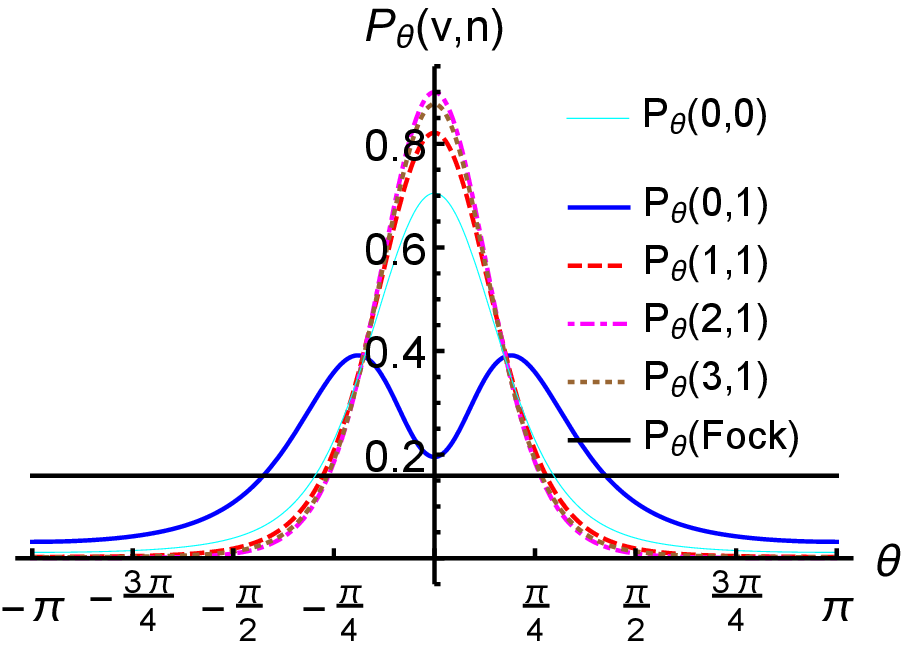}}
\quad{}\quad{}\subfigure[]{ \includegraphics[scale=0.5]{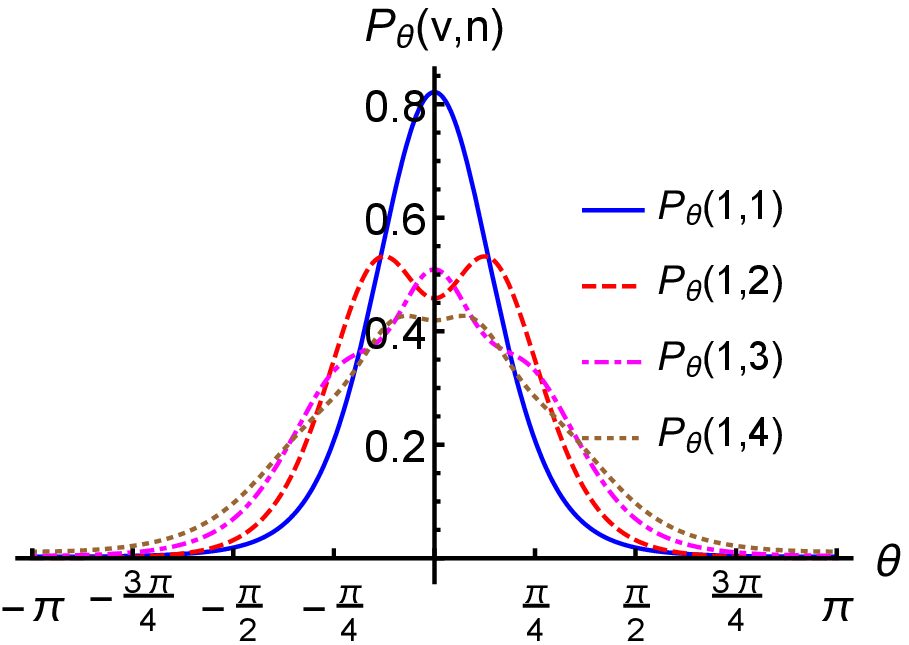}}\\
 \subfigure[]{\includegraphics[scale=0.4]{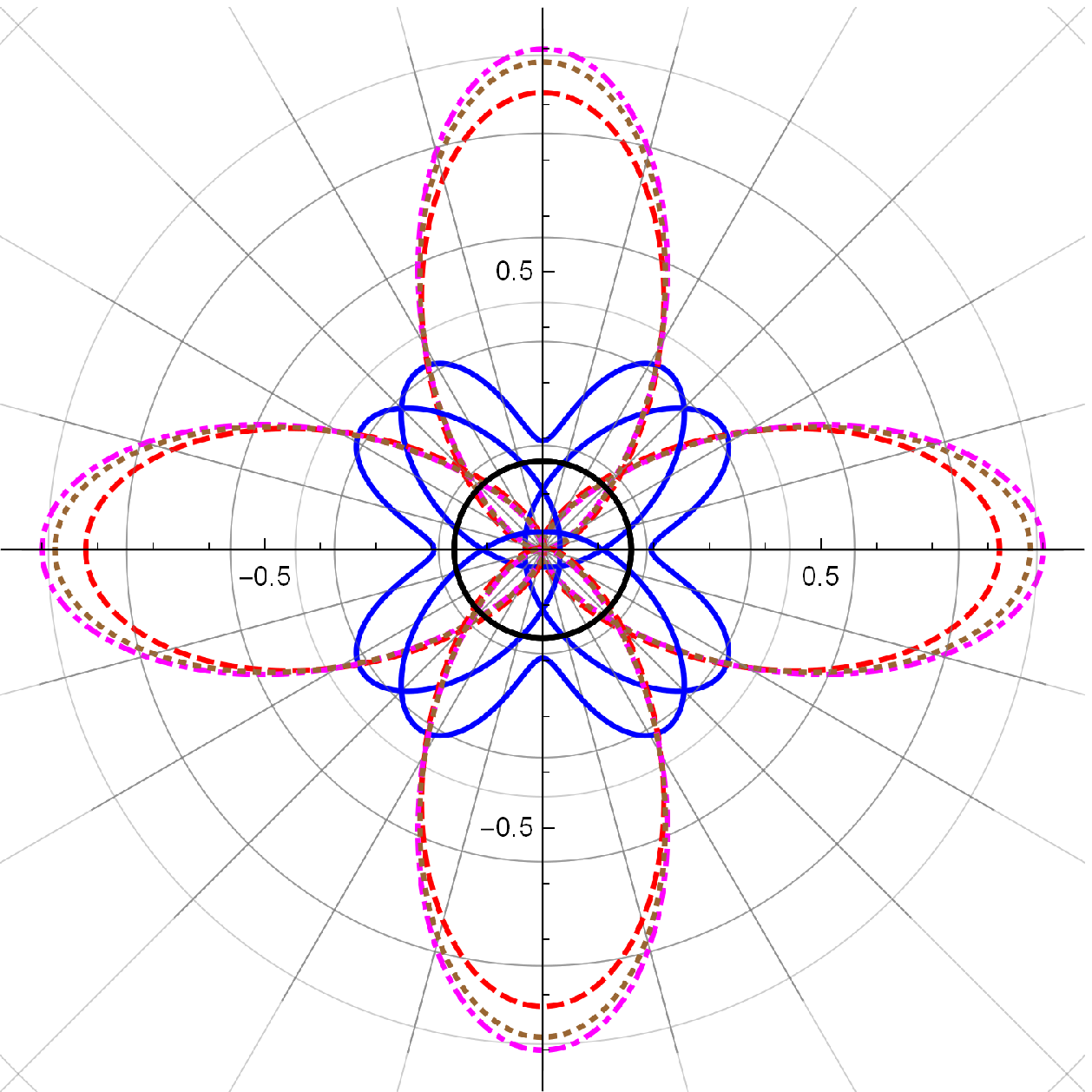}}\quad{}\quad{}\subfigure[]{
\includegraphics[scale=0.4]{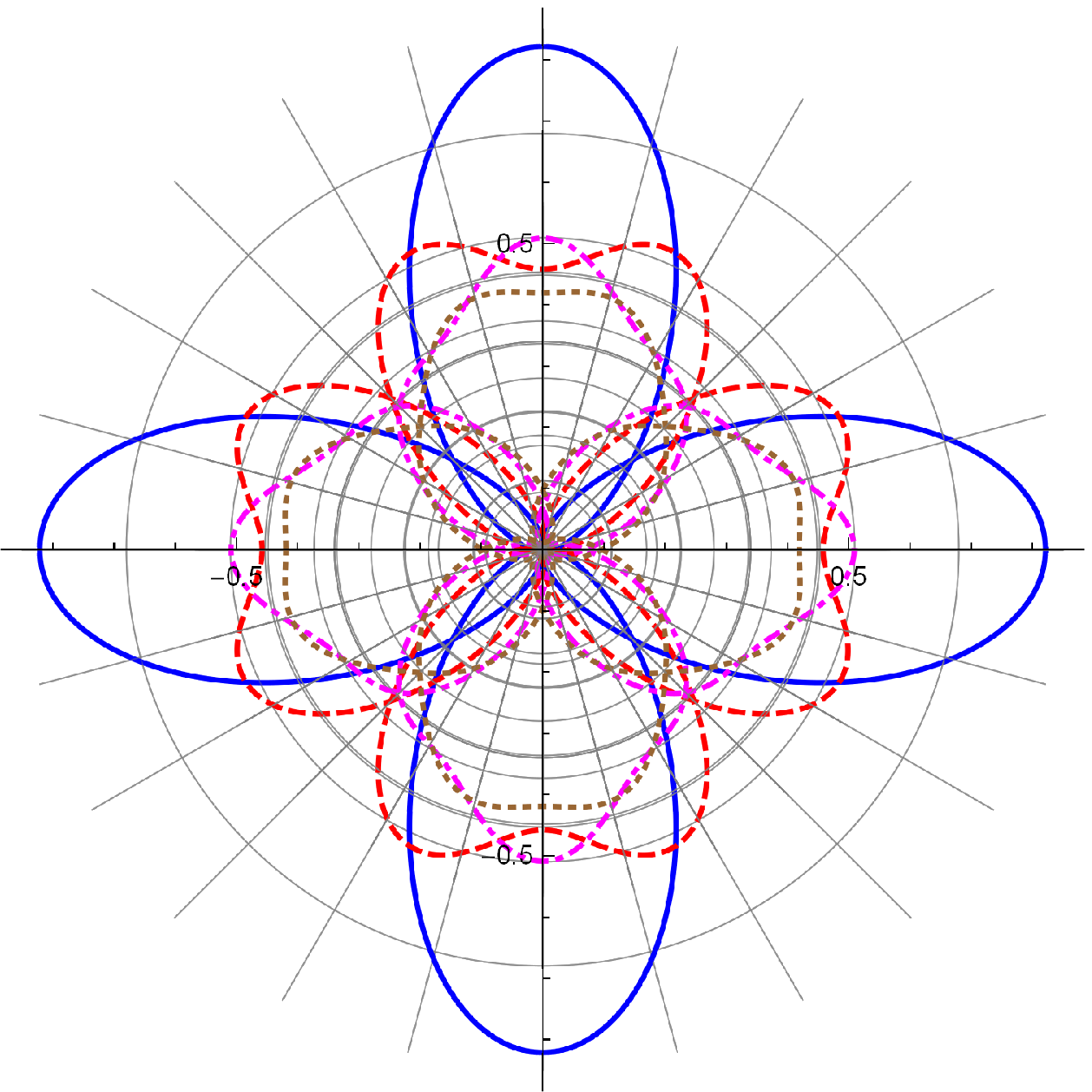}}
\caption{\label{fig:Phase-Distribution-Function-1} (Color online) Variation
of phase distribution function with phase parameter for PSDFS with
displacement parameter $\left|\alpha\right|=1$ for different values
of photon subtraction ((a) and (c)) and Fock parameters ((b) and (d)).
The phase distribution is shown using both two-dimensional ((a) and
(b) with $\theta_{2}=0$) and polar ((c) and (d)) plots. In (c) and
(d), $\theta_{2}=\frac{n\pi}{2}$ with integer $n\in\left[0,3\right]$,
and the legends are same as in (a) and (b), respectively.}
\end{figure}

To begin with, we compute the analytic expressions of $P_{\theta}$
for the PADFS and PSDFS, using Eq. (\ref{eq:Phase-Distridution})
as

\begin{equation}
\begin{array}{lcl}
P_{\theta}\left(u,n\right) & = & \frac{1}{2\pi}\dfrac{\left|N_{+}\right|^{2}}{n!}\sum\limits _{p,p^{\prime}=0}^{n}{n \choose p}{n \choose p^{\prime}}\exp\left[-\mid\alpha\mid^{2}\right]\left|\alpha\right|^{2n-p-p^{\prime}}\\
 & \times & \sum\limits _{m,m^{\prime}=0}^{\infty}\frac{(-\left|\alpha\right|)^{m+m^{\prime}}\sqrt{(m+p+u)!(m^{\prime}+p^{\prime}+u)!}}{m!m^{\prime}!}\exp[\iota\left(\theta-\theta_{2}\right)(m^{\prime}+p^{\prime}-m-p)]
\end{array}\label{eq:PA-phase}
\end{equation}
and

\begin{equation}
\begin{array}{ccc}
P_{\theta}\left(v,n\right) & = & \frac{1}{2\pi}\dfrac{\left|N_{-}\right|^{2}}{n!}\sum\limits _{p,p^{\prime}=0}^{n}{n \choose p}{n \choose p^{\prime}}\exp\left[-\mid\alpha\mid^{2}\right]\left|\alpha\right|^{2n-p-p^{\prime}}\\
 & \times & \sum\limits _{m,m^{\prime}=0}^{\infty}\frac{(-\left|\alpha\right|)^{m+m^{\prime}}(m+p)!(m^{\prime}+p^{\prime})!}{m!m^{\prime}!\sqrt{(m+p-v)!(m^{\prime}+p^{\prime}-v)!}}\exp[\iota\left(\theta-\theta_{2}\right)(m^{\prime}+p^{\prime}-m-p)],
\end{array}\label{eq:PS-phase}
\end{equation}
respectively. Since the obtained expressions in Eqs. (\ref{eq:PA-phase})
and (\ref{eq:PS-phase}) are complex in nature, we depict numerical
(graphical) analysis of the obtained results in Figs. \ref{fig:Phase-Distribution-Function}
and \ref{fig:Phase-Distribution-Function-1} for PADFS and PSDFS,
respectively. Specifically, in Fig. \ref{fig:Phase-Distribution-Function}
(a), we have shown the variation of phase distribution with phase
parameter $\theta$ for different number of photon added in the displaced
single photon Fock state ($D\left(\alpha\right)\left|1\right\rangle $)
for $\theta_{2}=0$. A uniform phase distribution for Fock state (with
a constant value of $\frac{1}{2\pi}$) is found to transform to one
that decreases for higher values of phase and possess a dip in the
phase distribution for $\theta=0$, which can be thought of as an
approach to the Fock state. In fact, in case of classical states,
$P_{\theta}$ has a peak at zero phase difference $\theta-\theta_{2}$,
and therefore, this contrasting behavior can be viewed as signature
of quantumness of DFS. However, with the increase in the number of
photons added to the DFS, the phase distribution of the PADFS is observed
to become narrower. In fact, a similar behavior with increase in the mean photon number of coherent state was observed
previously \cite{agarwal1992classical}. It is imperative to state
that $P_{\theta}$ in case of higher number of photon added to DFS
has similar but narrower distribution than that of coherent state.
In contrast, with increase in the Fock parameter, the phase distribution
is observed to become broader (cf. Fig. \ref{fig:Phase-Distribution-Function}
(b)). Thus, the increase in the number of photons added and the increase
in Fock parameter have opposite effects on the phase distribution.
The same is also illustrated through the polar plots in Fig. \ref{fig:Phase-Distribution-Function}
(c)-(d), which not only reestablish the same fact,
but also illustrate the dependence of $P_{\theta}$ on the phase
of the displacement parameter. Specifically, the obtained phase distribution
remains symmetric along the value of phase $\theta_{2}$ (i.e., $P_{\theta}$
is observed to have a mirror symmetry along $\theta=\theta_{2}$)
of the displacement parameter. The phase distribution of Fock state
is shown by a black circle in the polar plot.

Instead of photon addition, if we subtract photons from the DFS, a
similar effect on the phase distribution to that of photon addition
is observed. Further, a comparison between photon addition and subtraction
on the phase distribution establishes that a single photon subtraction
has a prominent impact on phase distribution when compared to that
of single photon addition, i.e., the distribution can be observed
to be narrower than that of coherent state in most of the cases for
$u=v$. For instance, single photon added (subtracted) DFS is broader
(narrower) than corresponding coherent state. Similarly, with the
increase in the value of Fock parameter, we can observe more changes
on PSDFS than what was observed in PADFS, i.e., the phase distribution
broadens more with Fock parameter for PSDFS. Note that $P_{\theta}$
has a peak at $\theta=\theta_{2}$ only for photon addition $u>n$,
while in case of photon subtraction it can be observed for $v\geq n$.
With the increase in the amplitude of displacement parameter ($\left|\alpha\right|$)
initially the phase distribution becomes narrower, which is further
supported by both addition and subtraction of photons, but it becomes
broader again for very high $\left|\alpha\right|$ (figure is not
shown here).

\subsection{Angular $Q$ function of PADFS and PSDFS}

\begin{figure}
\centering{}

\subfigure[]{\includegraphics[scale=0.4]{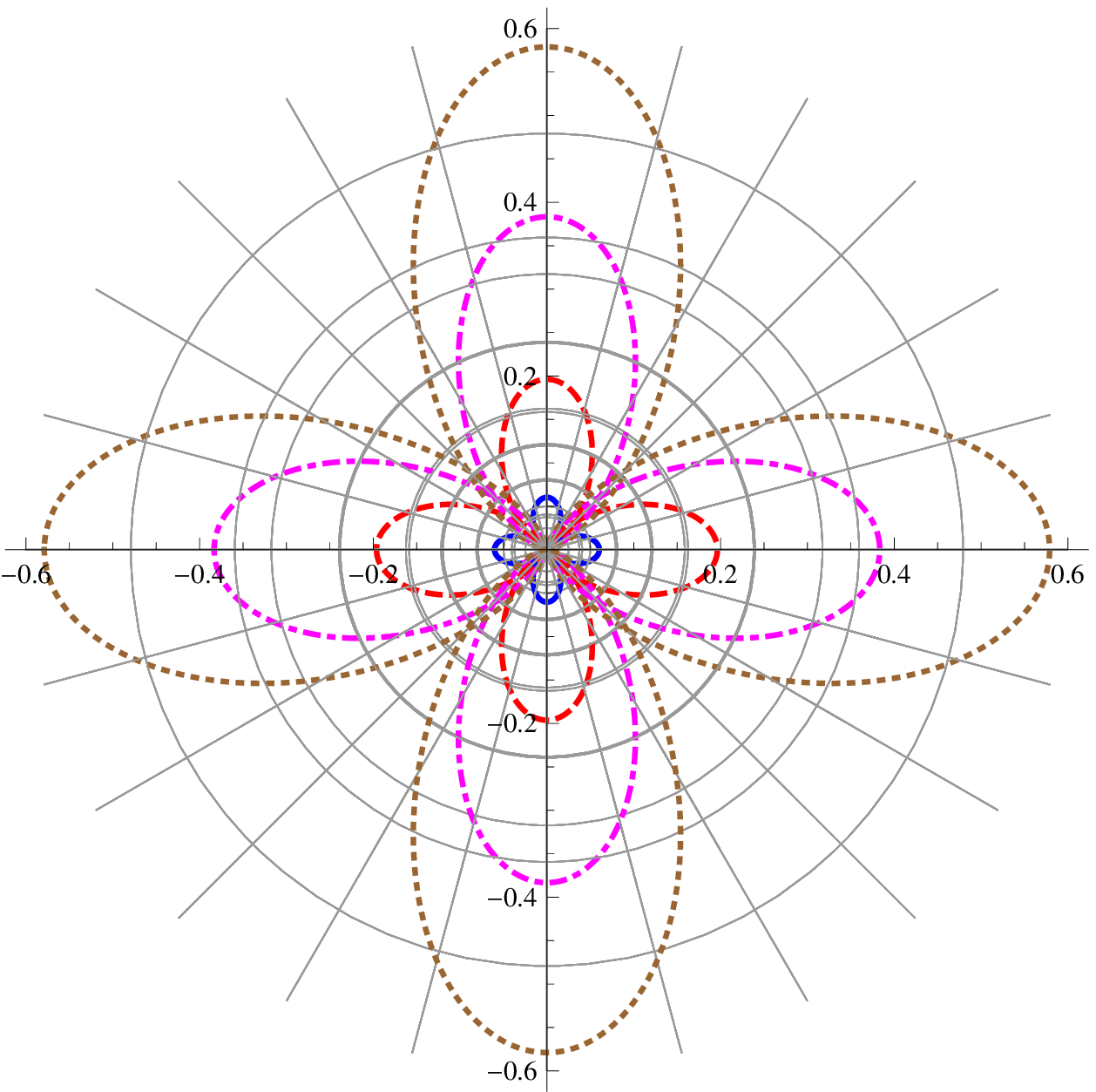}}
\quad{}\quad{}\subfigure[]{ \includegraphics[scale=0.4]{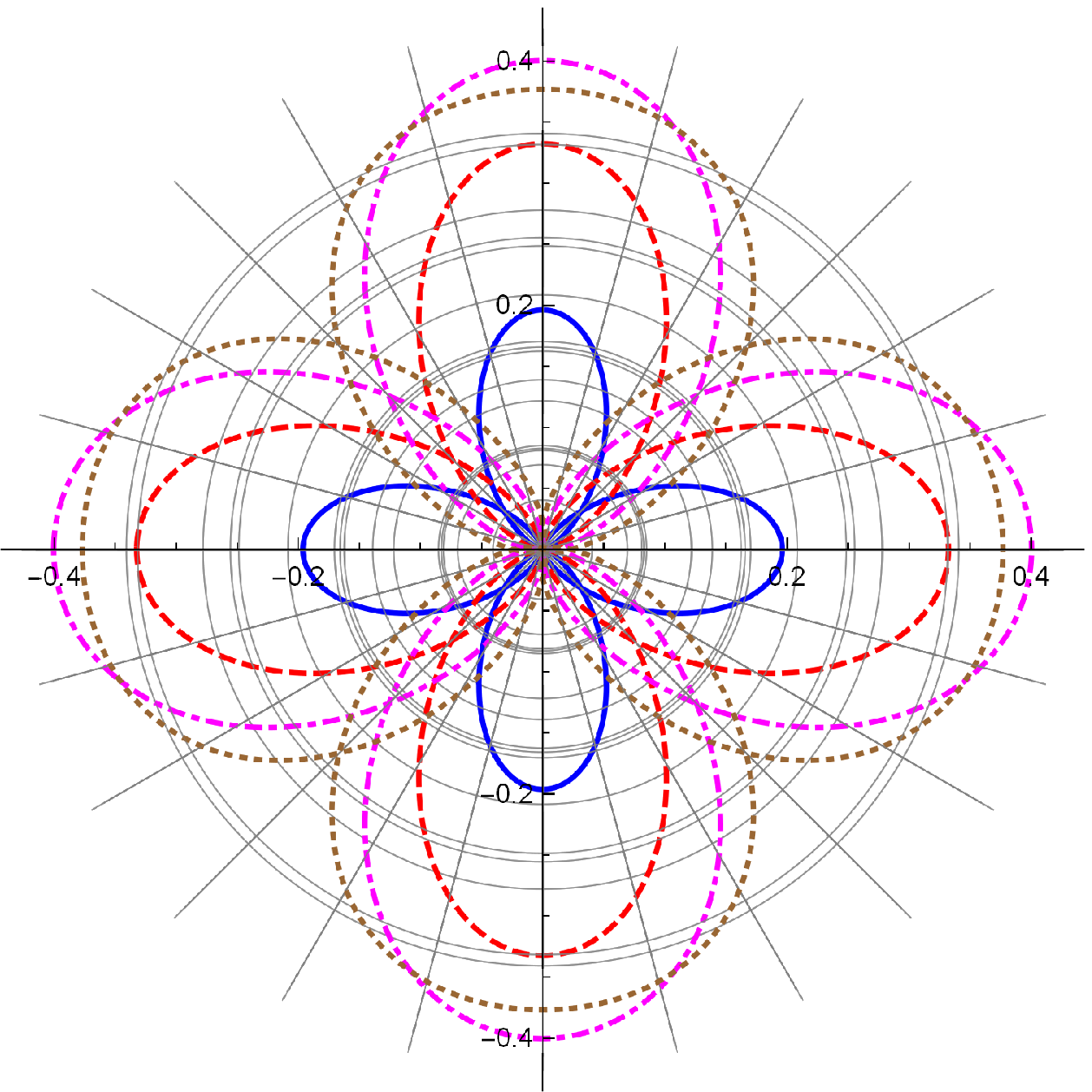}}\\
 \subfigure[]{\includegraphics[scale=0.4]{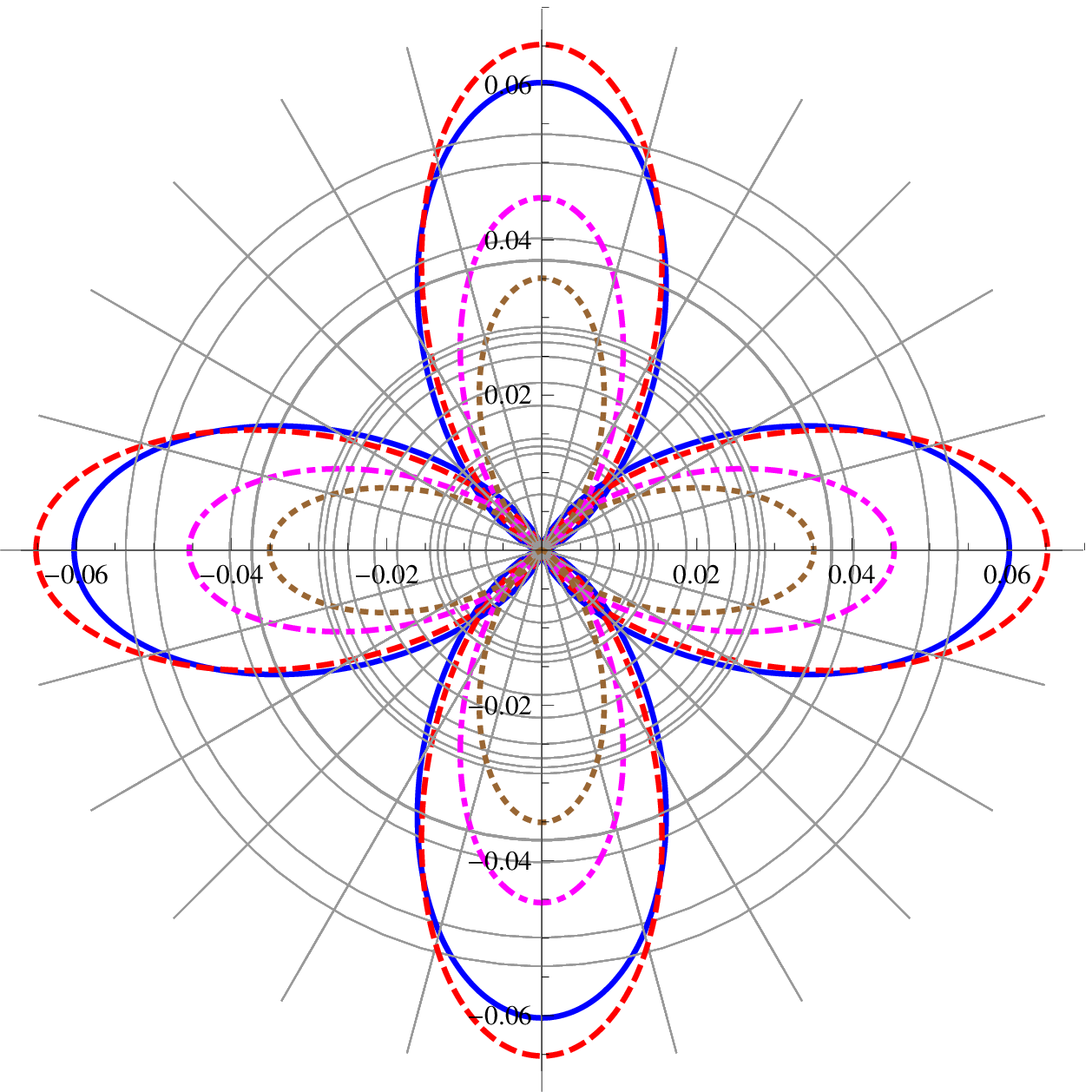}}\quad{}\quad{}\subfigure[]{
\includegraphics[scale=0.4]{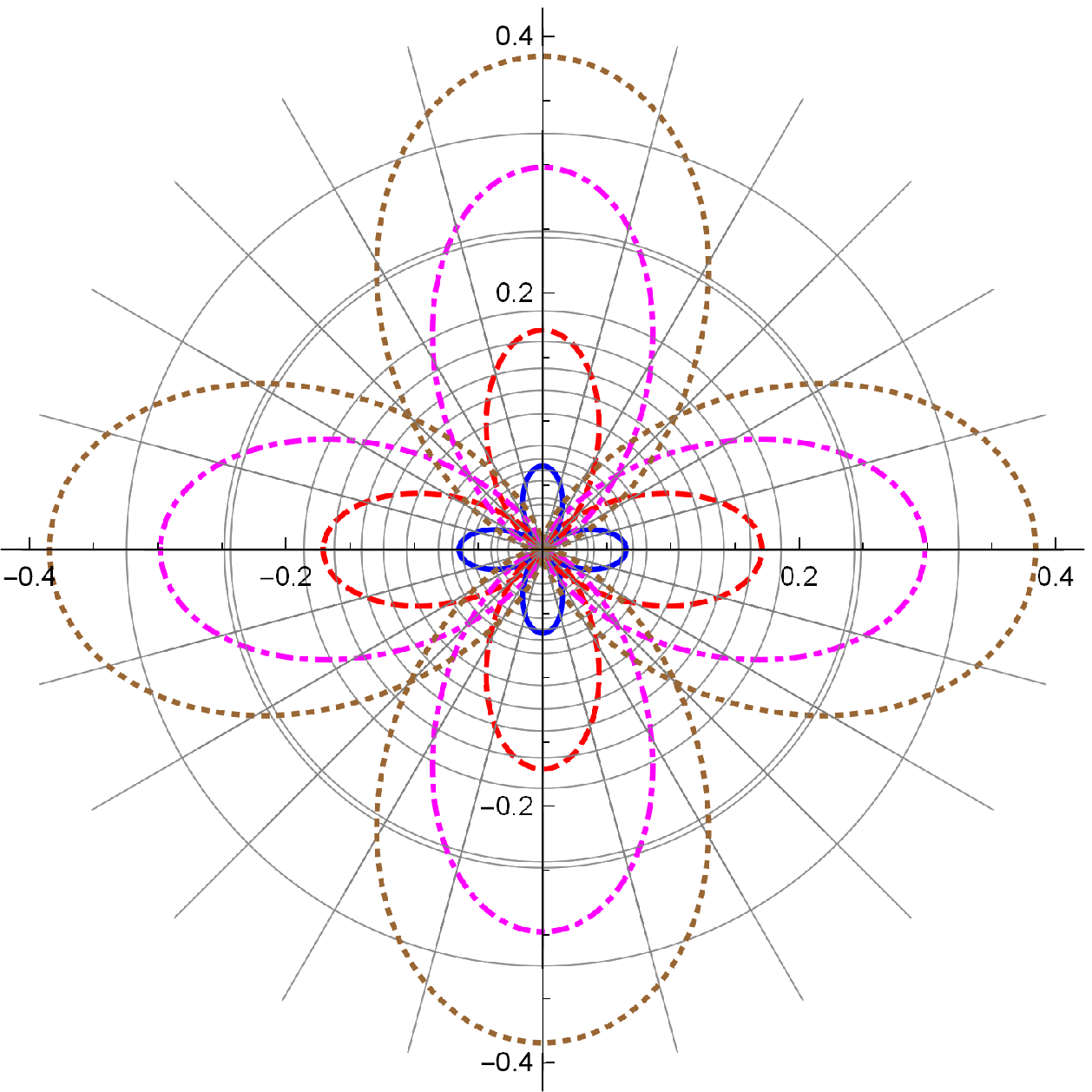}}\caption{\label{fig:Angular Q function} (Color online) The polar plots for
angular $Q$ function for PADFS (in (a) and (b)) and PSDFS (in (c)
and (d)) for displacement parameter $\left|\alpha\right|{\rm =}1$
and $\theta_{2}=\frac{n\pi}{2}$ with integer $n\in\left[0,3\right]$
for different values of photon addition/subtraction and Fock parameters.
In (a) and (c), for $n=1$, the smooth (blue), dashed (red), dot-dashed
(magenta), and dotted (brown) lines correspond to photon addition/subtraction
0, 1, 2, and 3, respectively. In (b) and (d), for the single photon
added/subtracted displaced Fock state, the smooth (blue), dashed (red),
dot-dashed (magenta), and dotted (brown) lines correspond to Fock
parameter 1, 2, 3, and 4, respectively. }
\end{figure}

We further discuss a phase distribution based on $Q$ function using
Eq. (\ref{eq:ang-Qf}). In this particular case, we have obtained
the angular $Q$ function from the $Q$ functions of the PADFS and
PSDFS reported as Eqs. (15)-(16) in \cite{malpani2019lower}. Specifically,
we have shown the effect of photon addition on the DFS ($D\left(\alpha\right)\left|1\right\rangle $)
for a specific value of the displacement parameter in Fig. \ref{fig:Angular Q function}
(a) for angular $Q$ function. One can clearly see that the polar
plots show an increase in the peak (located at $\theta_{1}=\theta_{2}$)
of the distribution with photon addition. Further, one can compare
the behavior of $Q_{\theta_{1}}$ with $P_{\theta}$ in Fig. \ref{fig:Phase-Distribution-Function}
and observe that they behave quite differently (as reported in \cite{agarwal1992classical}
for the coherent states), other than increase in the peak of the distribution.
Specifically, $P_{\theta}$ has a peak at $\theta=\theta_{2}$ only
for $u>n$, while $Q_{\theta_{1}}$ is always peaked at the phase
of the displacement parameter which also becomes a line of symmetry.
Interestingly, the effect of increase in the Fock parameter of PADFS
on $Q_{\theta_{1}}$is similar but less prominent in comparison to
photon addition. This is in quite contrast of that observed for $P_{\theta}$
(in Figs. \ref{fig:Phase-Distribution-Function} and \ref{fig:Angular Q function}
(b)). In case of PSDFS, both photon subtraction and Fock parameter
have completely different effects on $Q_{\theta_{1}}$ (cf. Fig. \ref{fig:Angular Q function}
(c)-(d)) which is also in contrast to that on corresponding $P_{\theta}$
(shown in Fig. \ref{fig:Phase-Distribution-Function-1}). Specifically,
with increase in photon subtraction the angular $Q$ function becomes
narrower peaked at $\theta=\theta_{2}$, but for larger number of
photon subtraction the peak value decreases quickly. However, with
increasing Fock parameter (cf. Fig. \ref{fig:Angular Q function}
(d)), $Q_{\theta_{1}}$ behaves much like photon addition on DFS (shown
in Fig. \ref{fig:Angular Q function} (a)). The observed behavior
shows the relevance of studying both these phase distributions due
to their independent characteristics. 

\subsection{Quantum phase fluctuation of PADFS and PSDFS}

\begin{figure}
\centering{}

\subfigure[]{\includegraphics[scale=0.5]{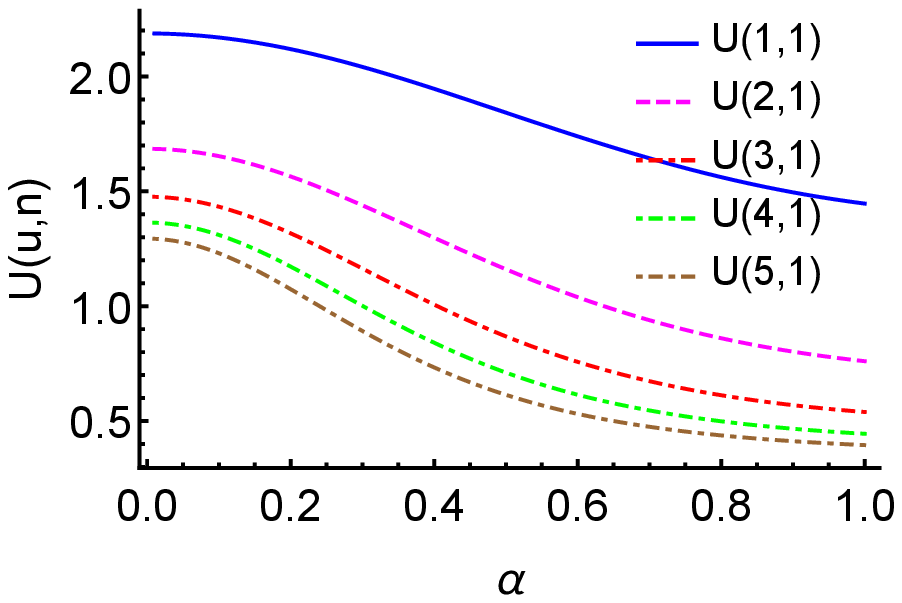}}
\quad{}\quad{}\subfigure[]{ \includegraphics[scale=0.5]{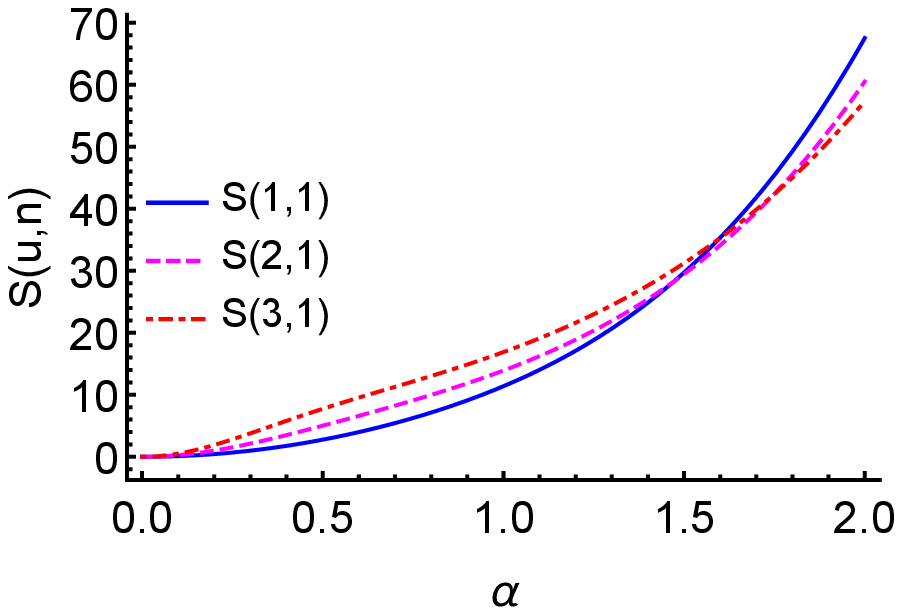}}
\quad{}\quad{}\subfigure[]{ \includegraphics[scale=0.5]{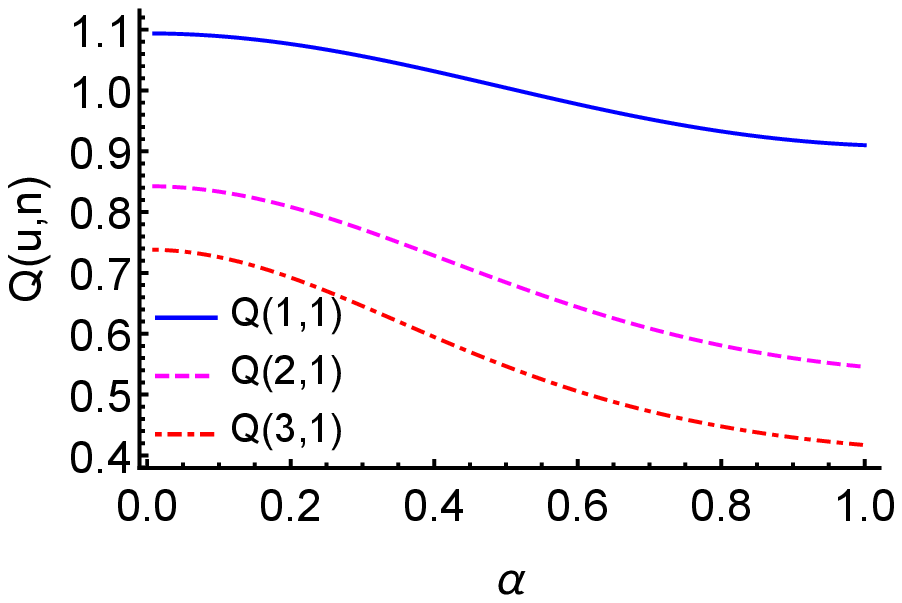}}\\
 \subfigure[]{\includegraphics[scale=0.5]{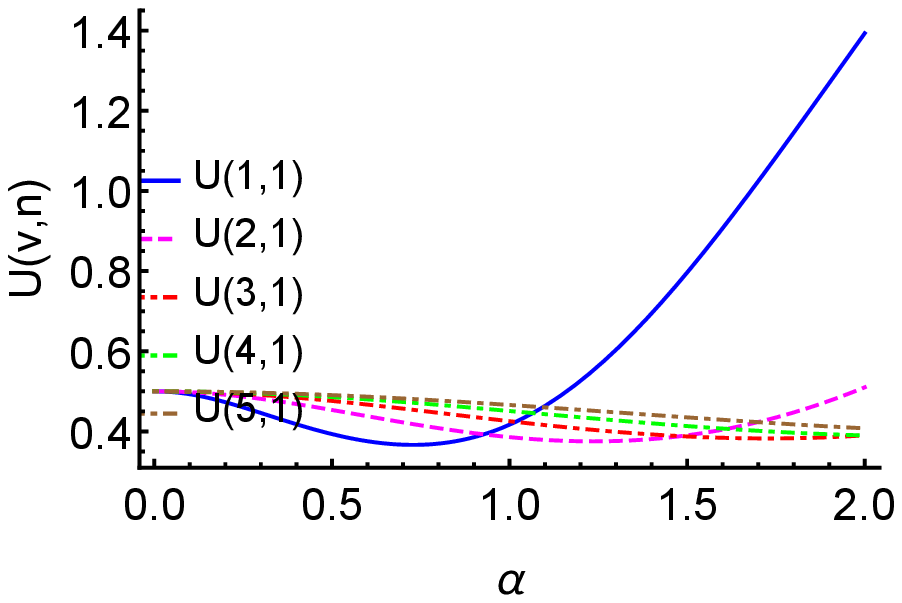}}\quad{}\quad{}\subfigure[]{
\includegraphics[scale=0.5]{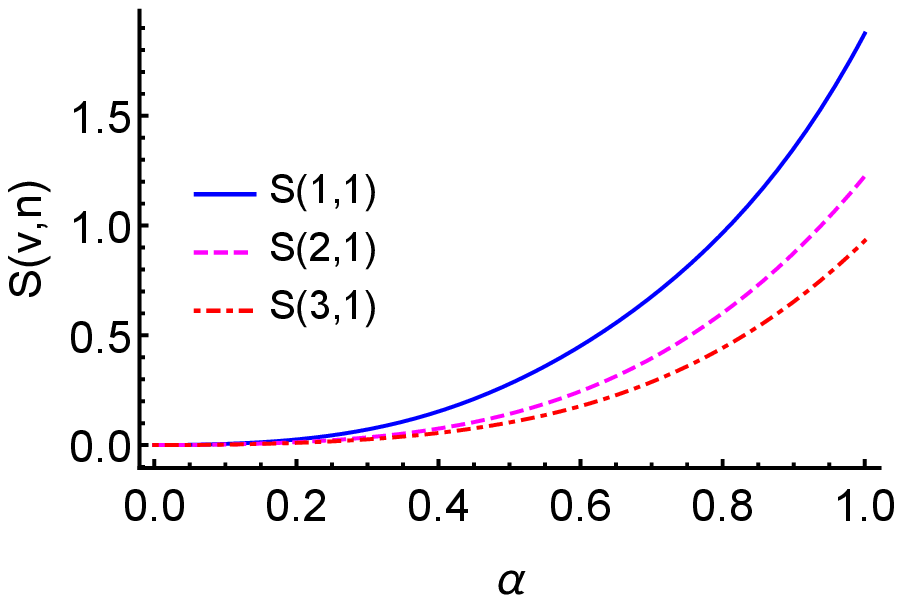}}\quad{}\quad{}\subfigure[]{
\includegraphics[scale=0.5]{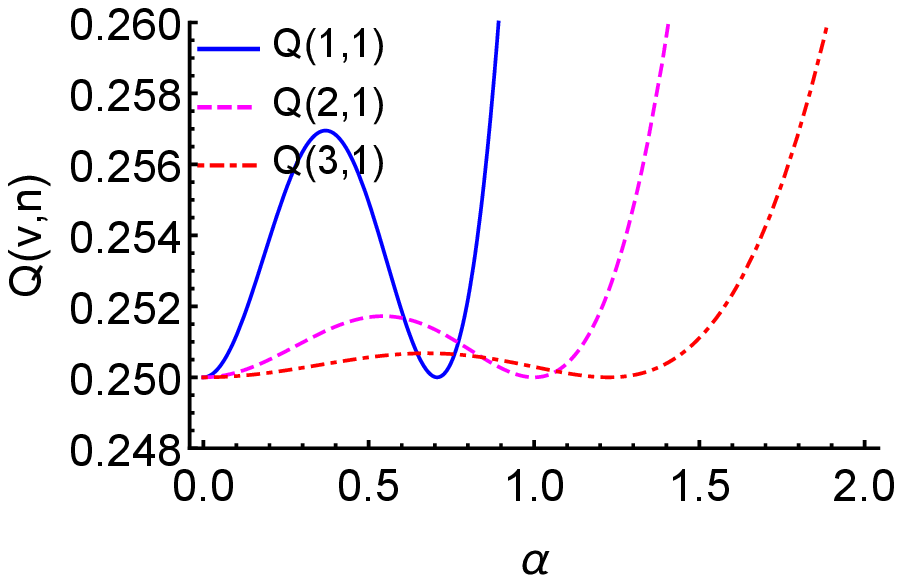}}
\caption{\label{fig:phase fluctuation} (Color online) Variation of three phase
fluctuation parameters introduced by Carruthers and Nieto with the
displacement parameter with $\theta_{2}=0$. The values of photon
addition ($u$), subtraction ($v$), and Fock parameter $n=1$ are
given in the legends. Parameter $U\left(i,n\right)\,\forall i\in\left\{ u,v\right\} $
also illustrates antibunching in the states for values less than $\frac{1}{2}$. }
\end{figure}

Carruthers and Nieto \cite{carruthers1968phase} introduced three
parameters to study quantum phase fluctuation (\ref{eq:fluctuation3})-(\ref{eq:fluctuation5}).
It was only in the recent past that some of the present authors provided
a physical meaning to one of these parameters by establishing its
relation with antibunching and sub-Poissonian photon statistics. Thus,
the quantum phase fluctuation studied here using three parameters
will also be used to witness the nonclassical nature of the quantum
states under consideration. Here, the effect of photon addition/subtraction
and displacement parameters on these fluctuation parameters is also
studied (shown in Fig. \ref{fig:phase fluctuation}). Specifically,
Fig. \ref{fig:phase fluctuation} (a)-(c) show variation of the three parameters of quantum phase fluctuation for different values
of the number of photons added in the displaced Fock state ($D\left(\alpha\right)\left|1\right\rangle $)
with displacement parameter $\left|\alpha\right|$. It may be clearly
observed that two of the quantum phase fluctuation parameters, namely
$U\left(u,n\right)$ and $Q\left(u,n\right)$ decrease with the value
of displacement parameter, while $S\left(u,n\right)$ increases with
$\left|\alpha\right|$. Interestingly, the photon
addition and increase in the displacement parameter exhibit the same
effect on all three quantum phase fluctuation parameters for PADFS,
while for higher values of displacement parameter $S\left(u,n\right)$
show completely opposite effect of photon addition. In contrast, $U\left(v,n\right)$
for $v$ subtracted photons from $D\left(\alpha\right)\left|1\right\rangle $
is found to increase (decrease) with photon subtraction while decrease
(increase) with the displacement parameter for small (large) value
of $\left|\alpha\right|$ (cf. Fig. \ref{fig:phase fluctuation} (d)).
On the other hand, parameter $S\left(v,n\right)$
is also observed to increase (decrease) with $\left|\alpha\right|$
($v$) as shown in Fig. \ref{fig:phase fluctuation} (e). The third
parameter $Q\left(v,n\right)$ shows slightly complex behavior for
PSDFS with both $\left|\alpha\right|$ and $v$ (cf. Fig. \ref{fig:phase fluctuation}
(f)) as it behaves analogous to PADFS for each subtracted photon for
both small and large values of the displacement parameter (when it
increases with $\left|\alpha\right|$), but for intermediate values
the behavior is found to be completely opposite. 

As  mentioned previously, $U\left(i,n\right)\,\forall i\in\left\{ u,v\right\} $
has a physical significance as a witness of antibunching for values
of this parameter less than $\frac{1}{2}$, Fig. \ref{fig:phase fluctuation}
(a) and (d) can be used to perform similar studies for PADFS and PSDFS,
respectively. In case of PADFS, we can observe this relevant parameter
to become less than $\frac{1}{2}$, and thus to illustrate the presence
of antibunching, only at higher values of the displacement parameter
and photon added to the displaced Fock state. In contrast, PSDFS shows
the presence of this nonclassical feature in all cases.Thus, occurrence of antibunching in PADFS and PSDFS is established
here through this phase fluctuation parameter. Interestingly, a similar
dependence of antibunching in PADFS and PSDFS has been recently reported
by us \cite{malpani2019lower} using a different criterion. Further,
one can observe from the expression of $U$ in Eq. (\ref{eq:fluctuation3})
that it is expected to be independent of the phase of the displacement parameter,
which can also be understood from the use of this parameter as a witness
for an intensity moments based nonclassical feature. In contrast,
$S$ and $Q$ in Eqs. (\ref{eq:fluctuation4})-(\ref{eq:fluctuation5})
show dependence on the phase of displacement parameter. Here, we have
not discussed the effect of Fock parameter in detail, but in case
of photon addition, $u$ and $n$ have same (opposite) effects on
$S$ ($U$ and $Q$) parameter(s). Fock parameter has always shown
opposite effect of photon subtraction on all three phase fluctuation
parameters, and thus nonclassicality revealed by $U$ can be enhanced
with Fock parameter. The relevance of Fock parameter can also be visualized
by observing the fact that the single photon subtracted coherent state
has $U=0.5$ (which is consistent with the value zero of the antibunching
witness reported in \cite{thapliyal2017comparison}). Thus, in this
case, the origin of the induced antibunching can be attributed to
the non-zero value of Fock parameter. 

\subsection{Phase Dispersion}

\begin{figure}
\centering{}

\subfigure[]{\includegraphics[scale=0.5]{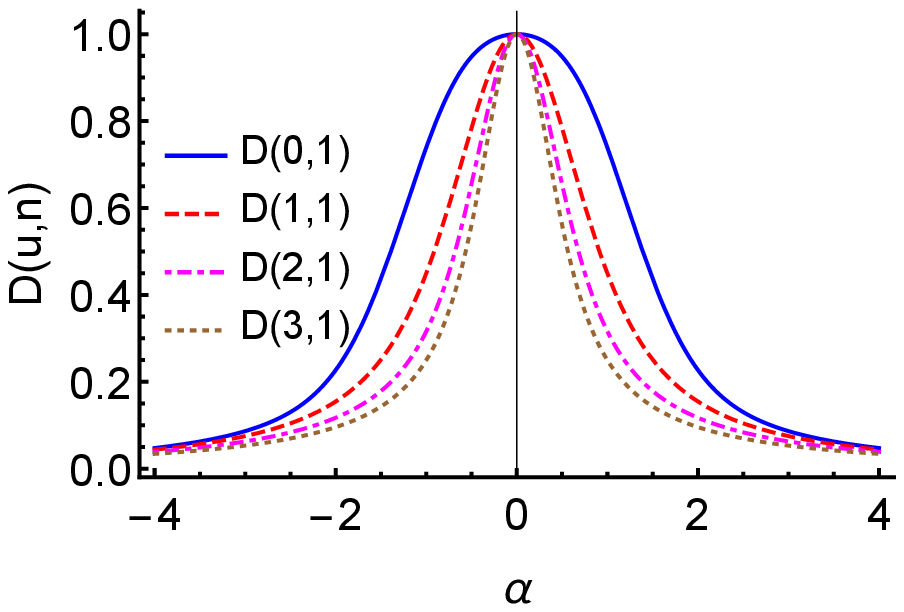}}
\quad{}\quad{}\subfigure[]{ \includegraphics[scale=0.5]{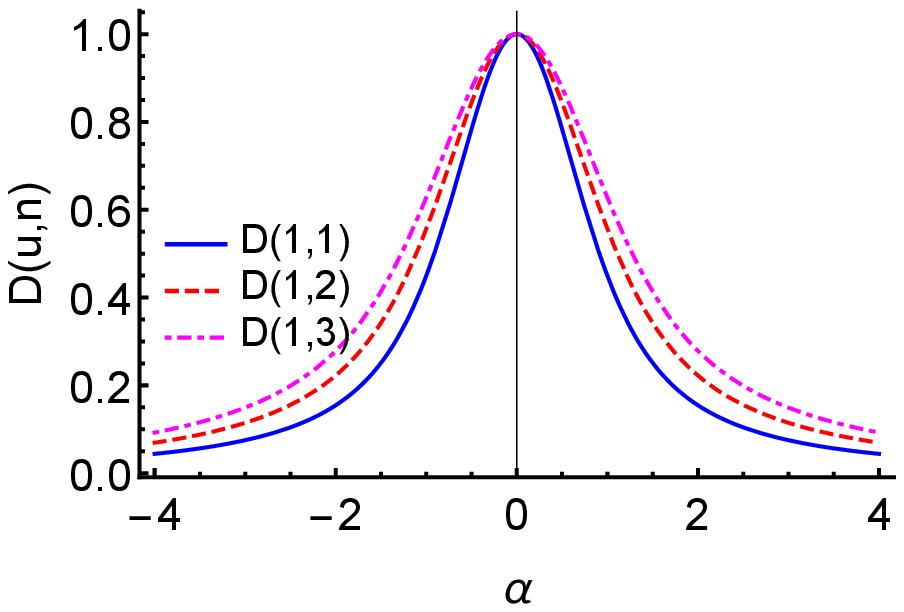}}\\
 \subfigure[]{\includegraphics[scale=0.5]{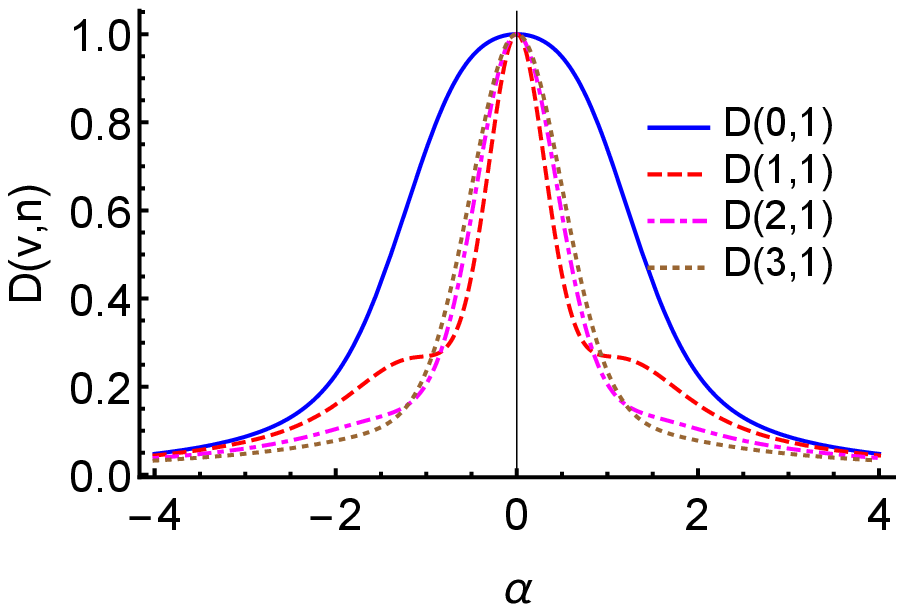}}\quad{}\quad{}\subfigure[]{
\includegraphics[scale=0.5]{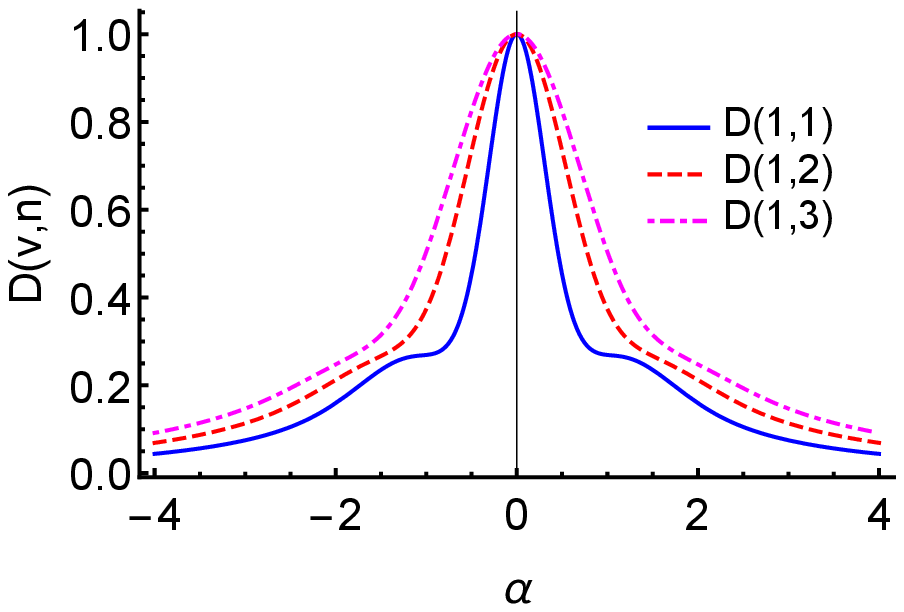}}
\caption{\label{fig:Phase-Dispersion} (Color online) Variation of phase dispersion
for PADFS (in (a) and (b)) and PSDFS (in (c) and (d)) with displacement
parameter for an arbitrary $\theta_{2}$. Dependence on different
values of photon added/subtracted and the initial Fock state $\left|1\right\rangle $
(in (a) and (c)), while on different values of Fock parameter for
single photon added/subtracted state (in (b) and (d)).}
\end{figure}

We compute a measure of quantum phase fluctuation based on quantum
phase distribution, the phase dispersion (\ref{eq:Dispersion}), for
both PADFS and PSDFS to perform a comparative study between them.
Specifically, the maximum value of dispersion is 1 which corresponds
to the uniform phase distribution, i.e., $P_{\theta}=\frac{1}{2\pi}$.
Both PADFS and PSDFS show a uniform distribution for the displacement
parameter $\alpha=0$ (cf. Fig. \ref{fig:Phase-Dispersion}). It is
a justified result as both the states reduce to the Fock state in
this case. However, with the increase in the value of displacement
parameter quantum phase dispersion is found to decrease. This may
be attributed to the number-phase complimentarity \cite{banerjee2010complementarity,srikanth2009complementarity,srikanth2010complementarity},
which leads to smaller phase fluctuation with increasing variance
in the number operator at higher values of displacement parameter.
Thus, with an increase in the average photon number by increasing
the displacement parameter, phase dispersion decreases for both PADFS
and PSDFS. Addition of photons in DFS leads to decrease in the value
of phase dispersion, while subtraction of photons has more complex
effect on phase dispersion (cf. Fig. \ref{fig:Phase-Dispersion} (a)
and (c)). Specifically, for the smaller values of the displacement
parameter ($\left|\alpha\right|<1$), the phase dispersion parameter
behaves differently for $v\leq n$ and $v>n$. This can be attributed
to the sub-Poissonian photon statistics for $v\leq n$ with $\left|\alpha\right|<1$
as well as the small value of average photon number (Fig. \ref{fig:phase fluctuation}
(d)). However, at the higher values of the displacement parameter
$D$ for the PSDFS behaves in a manner analogous to the PADFS. Interestingly,
increase in the Fock parameter shows similar effect on PADFS and PSDFS
in Fig. \ref{fig:Phase-Dispersion} (b) and (d), respectively.

\subsection{Phase sensing uncertainity for PADFS and PSDFS}

We finally discuss quantum phase estimation using Eq. (\ref{eq:PE}),
assuming the two mode input state in the Mach-Zehnder interferometer
as $|\psi_{i}(j,n,\alpha)\rangle\otimes|0\rangle$. 
The expressions for the variance of
the difference in the photon numbers in the two output modes of the
Mach-Zehnder interferometer for input PADFS and PSDFS and the rest of the parameters required to study phase sensing are reported in Appendix. 

\begin{figure}
\centering{}

\subfigure[]{\includegraphics[scale=0.5]{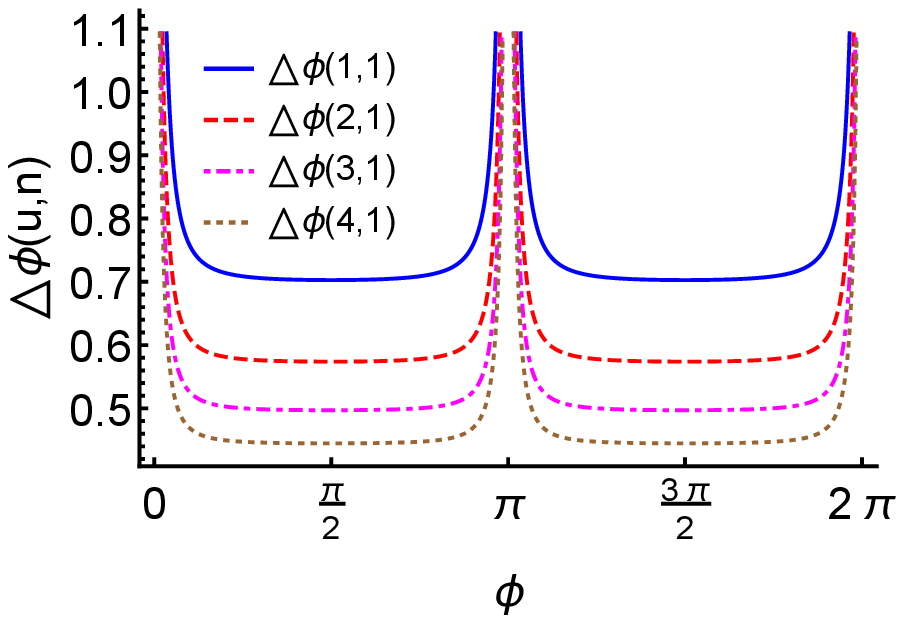}}
\quad{}\quad{}\subfigure[]{ \includegraphics[scale=0.5]{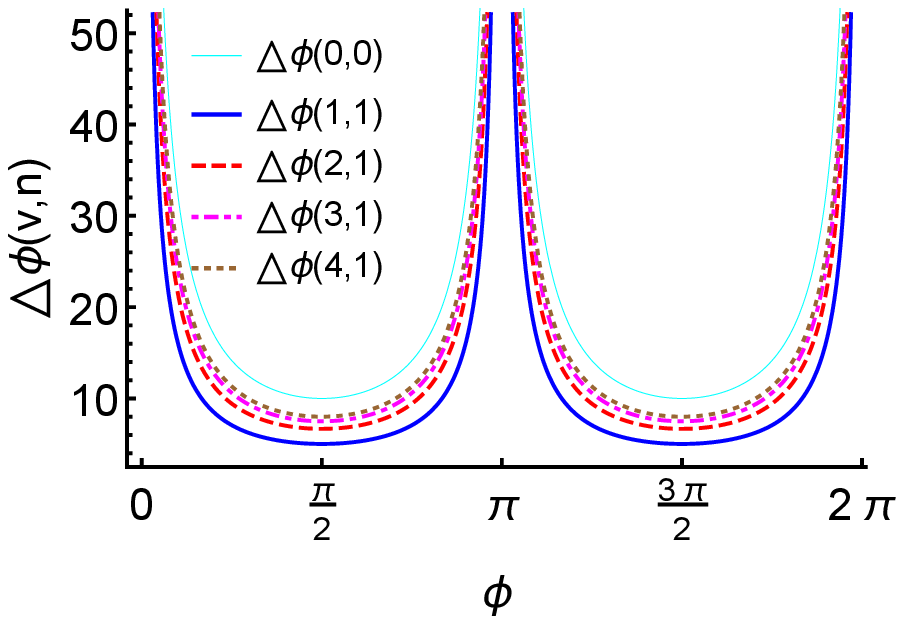}}\\
 \subfigure[]{\includegraphics[scale=0.5]{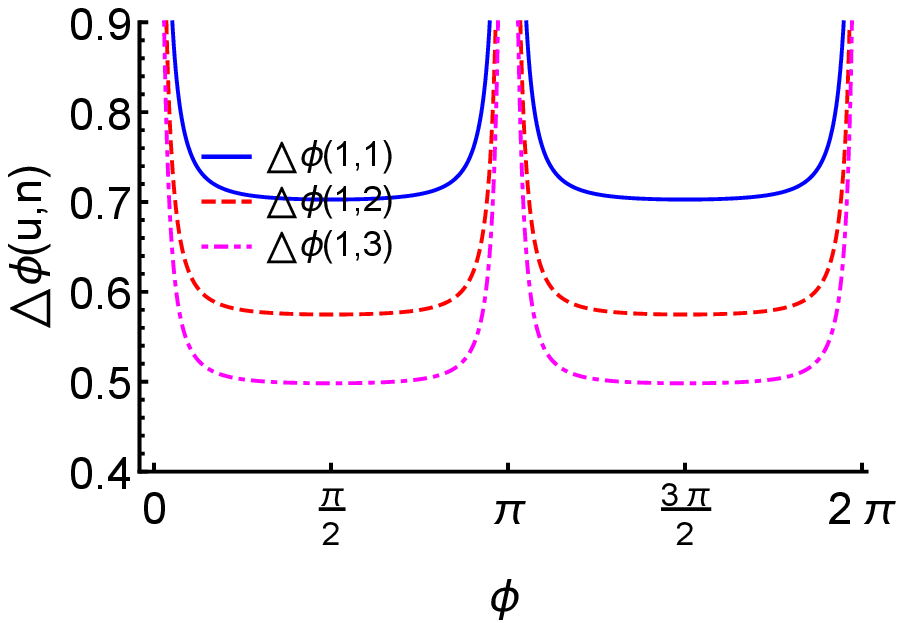}}\quad{}\quad{}\subfigure[]{
\includegraphics[scale=0.5]{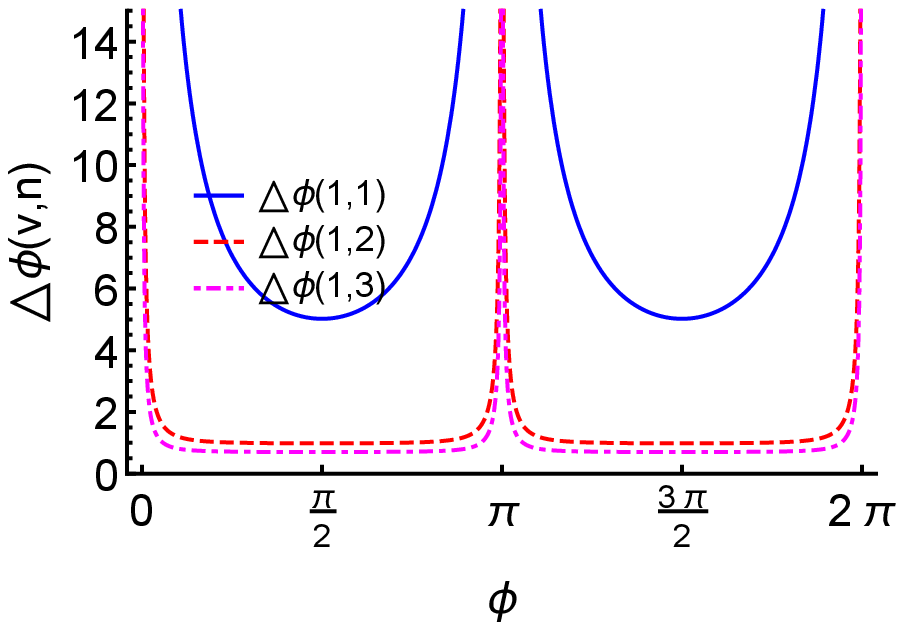}}
\caption{\label{fig:Phase sensing uncertainity}Phase sensing uncertainty for
(a) PADFS and (b) PSDFS as a function of phase to be estimated $\phi$
for different number of photon addition/subtraction with $n=1$. The
dependence for (c) PADFS and (d) PSDFS is also shown for different values of Fock parameters with
$u=1$ and $v=1$, respectively. In all cases, we have chosen $\alpha=0.1$.}
\end{figure}

The obtained expressions allow us to study the optimum
choice of state parameters for quantum phase estimation using PADFS and PSDFS. The variation of these parameters is shown
in Fig. \ref{fig:Phase sensing uncertainity}. Specifically, we have
shown that PSDFS is preferable over coherent state for phase estimation
(cf. Fig. \ref{fig:Phase sensing uncertainity} (b)). However, with
the increase in the photon subtraction this phase uncertainty parameter
is found to increase although remaining
less than corresponding coherent state value. In contrast, with photon
addition, advantage in phase estimation can be attained as the reduction
of the phase uncertainty parameter allows one to perform more precise
measurement. This advantage can be enhanced further by choosing large
values of photon addition and Fock parameter (cf. Fig. \ref{fig:Phase sensing uncertainity}
(a) and (c)). In a similar sense, appropriate choice of Fock parameter
would also be advantageous in phase estimation with
PSDFS as it decreases the phase uncertainty parameter, but still PADFS
remains preferable over PSDFS. This can further be controlled by an increase in $\left|\alpha\right|$ which decreases (increases) phase uncertainty parameter for PADFS (PSDFS).

\section{Conclusions \label{sec:Conclusions}}

A set of engineered quantum states can be obtained as the limiting
cases from the PADFS and PSDFS, i.e., DFS, coherent
state, photon added/subtracted coherent state, and Fock state. Specifically,
PADFS/PSDFS are obtained by application of unitary (displacement)
and non-unitary (addition and subtraction of photons) operations on
Fock state. In view of the fact that the Fock states have uniform
phase distribution, the set of unitary and non-unitary quantum state
engineering operations are expected to affect the phase properties
of the generated state. Therefore, here we have calculated quantum
phase distribution, which further helped in quantifying phase fluctuation
as phase dispersion. We have also computed the phase distribution
as the angular $Q$ function. We have further studied phase fluctuation
using three Carruthers and Nieto parameters, and have used one of
them to reveal the existence of antibunching in the quantum states
of our interest. 

Both the phase distribution and angular $Q$ functions are found to
be symmetric along the value of the phase of the displacement
parameter. The phase distribution is observed to become narrow and
peak(s) to increase with the amplitude of the displacement parameter
($\left|\alpha\right|$), which further becomes broader for higher
values of $\left|\alpha\right|$. Further, photon addition/subtraction
and Fock parameters are observed to have opposite effects on phase
distribution, i.e., distribution function becomes narrower (broader)
with photon addition/subtraction (Fock parameter). Among photon addition
and subtraction operations, subtracting a photon alters the phase
properties more than that of photon addition. Specifically, at the
small values of the displacement parameter ($\left|\alpha\right|<1$),
the phase properties of PSDFS for $v\leq n$ and $v>n$ behave differently.
This can be attributed to the fact that for $v\leq n$ with $\left|\alpha\right|<1$,
the average photon number becomes very small. Further, the peak of
the phase distribution remains at the phase of displacement parameter
only when the number of photons added/subtracted is more than that
of the Fock parameter. However, in case the number of photons subtracted
(added) is same as the Fock parameter, the peak of the phase distribution
is observed (not observed) at the phase of displacement parameter.
The angular $Q$ function can be observed to show similar dependence
on various parameters, but the peak of the distribution remains located
at the value of phase of the displacement parameter. The three phase
fluctuation parameters introduced by Carruthers and Nieto \cite{carruthers1968phase}
show phase properties of PADFS and PSDFS, while one of them, $U$
parameter also reveals antibunching in both PADFS and PSDFS. In this case, the
role of Fock parameter as antibunching inducing operation in PSDFS
is also discussed. Phase dispersion quantifying phase fluctuation
remains unity for Fock state reflecting uniform distribution, which
can be observed to decrease with increasing displacement parameter.
This may be attributed to the number-phase complimentarity as the
higher values of variance with increasing displacement parameter lead
to smaller phase fluctuation. Fock parameter and photon addition/subtraction
show opposite effects on the phase dispersion as it increases (decreases)
with $n$ ($u/v$). 

Finally, we have also discussed the advantage of the PADFS and PSDFS
in quantum phase estimation and obtained the set of optimized parameters
in the PADFS/PSDFS. Both photon addition and Fock parameter decrease
the uncertainty in phase estimation, while photon subtraction, though
performs better than coherent state is not as advantageous as $u$
or $n$. In \cite{ou1997fundamental}, it was established that signal-to-noise
ratio is significant only when the phase shift to measure is of the
same order as multiplicative inverse of the average photon number.
Therefore, in case of PADFS this limitation of quantum measurement
is expected to play an important role. Thus, we have shown here that
state engineering tools can be used efficiently to control the phase
properties of the designed quantum states for suitable applications.
The study can be performed for other such operations, like squeezing,
photon addition followed by subtraction or vice versa. 

We conclude the paper by noting that the method used for the present
study is quite general and can be applied to study phase properties
of many other engineered quantum states, and a major part of the
results presented here can be experimentally verified using the available technolgy. Consequently,
this work is expected to lead to a number of further studies and thus
to provide new insights into the phase properties of the quantum states.

\textbf{Acknowledgment:} KT acknowledges the financial support from
the Operational Programme Research, Development and Education - European
Regional Development Fund project no. CZ.02.1.01/0.0/0.0/16\_019/0000754
of the Ministry of Education, Youth and Sports of the Czech Republic.
AP and NA thank SERB, DST, India for the support provided through
the project number EMR/2015/000393. SB and VN thank CSIR, New Delhi
for support through the project No. 03(1369)/16/EMR-II. AP and KT also thank A Luks for his interest and technical remarks.

\bibliographystyle{apsrev4-1}
\bibliography{priya}

\appendix

\section*{Appendix: Phase estimation}

\setcounter{equation}{0} \renewcommand{\theequation}{A.\arabic{equation}}

The variance of
the difference in the photon numbers in the two output modes of the
Mach-Zehnder interferometer for input PADFS and PSDFS are

\begin{equation}
\begin{array}{lcl}
\left(\Delta{J_{z}}\right)_{+}^{2} & = & \cos^{2}\text{\ensuremath{\phi}}\left\{ \frac{1}{2}\frac{\left|N_{+}\right|^{2}}{n!}\sum\limits _{p,p'=0}^{n}{n \choose p}{n \choose p'}(-\alpha^{\star})^{(n-p)}(-\alpha)^{(n-p')}\exp\left[-\mid\alpha\mid^{2}\right]\right.\\
 & \times & \sum\limits _{m=0}^{\infty}\frac{\alpha^{m}(\alpha^{\star})^{m+p-p'}(m+p+u)!}{m!(m+p-p')!}\left[\frac{1}{2}\left(m+p+u\right)\left(m+p+u-1\right)+\left(m+p+u\right)-\frac{1}{2}\right.\\
 & \times & \frac{\left|N_{+}\right|^{2}}{n!}\sum\limits _{p,p'=0}^{n}{n \choose p}{n \choose p'}(-\alpha^{\star})^{(n-p)}(-\alpha)^{(n-p')}\exp\left[-\mid\alpha\mid^{2}\right] \left.\left.\sum\limits _{m=0}^{\infty}\frac{\alpha^{m}(\alpha^{\star})^{m+p-p'}(m+p+u)!\left(m+p+u\right)^{2}}{m!(m+p-p')!}\right]\right\} \\
 & + & \sin^{2}\text{\ensuremath{\phi}}\left\{ \frac{1}{4}\frac{\left|N_{+}\right|^{2}}{n!}\sum\limits _{p,p'=0}^{n}{n \choose p}{n \choose p'}(-\alpha^{\star})^{(n-p)}(-\alpha)^{(n-p')}\exp\left[-\mid\alpha\mid^{2}\right]\right. \left.\sum\limits _{m=0}^{\infty}\frac{\alpha^{m}(\alpha^{\star})^{m+p-p'}(m+p+u)!(m+p+u)}{m!(m+p-p')!}\right\} ,
\end{array}\label{eq:padfs-pe}
\end{equation}
and 
\begin{equation}
\begin{array}{lcl}
\left(\Delta{J_{z}}\right)_{-}^{2} & = & \cos^{2}\text{\ensuremath{\phi}}\left\{ \frac{1}{2}\frac{\left|N_{-}\right|^{2}}{n!}\sum\limits _{p,p'=0}^{n}{n \choose p}{n \choose p'}(-\alpha^{\star})^{(n-p)}(-\alpha)^{(n-p')}\exp\left[-\mid\alpha\mid^{2}\right]\right.\\
 & \times & \sum\limits _{m=0}^{\infty}\frac{\alpha^{m}(\alpha^{\star})^{m+p-p'}(m+p)!(m+p)!}{m!(m+p-p')!\left(m+p-v\right)!}\left[\frac{1}{2}\left(m+p-v\right)\left(m+p-v-1\right)+\left(m+p-v\right)-\frac{1}{2}\right.\\
 & \times & \frac{\left|N_{-}\right|^{2}}{n!}\sum\limits _{p,p'=0}^{n}{n \choose p}{n \choose p'}(-\alpha^{\star})^{(n-p)}(-\alpha)^{(n-p')}\exp\left[-\mid\alpha\mid^{2}\right] \left.\left.\sum\limits _{m=0}^{\infty}\frac{\alpha^{m}(\alpha^{\star})^{m+p-p'}(m+p)!\left(m+p-v\right)^{2}}{m!(m+p-p')!\left(m+p-v\right)!}\right]\right\} \\
 & + & \sin^{2}\text{\ensuremath{\phi}}\left\{ \frac{1}{4}\frac{\left|N_{-}\right|^{2}}{n!}\sum\limits _{p,p'=0}^{n}{n \choose p}{n \choose p'}(-\alpha^{\star})^{(n-p)}(-\alpha)^{(n-p')}\exp\left[-\mid\alpha\mid^{2}\right]\right. \left.\sum\limits _{m=0}^{\infty}\frac{\alpha^{m}(\alpha^{\star})^{m+p-p'}(m+p)!(m+p)!\left(m+p-v\right)}{m!(m+p-p')!\left(m+p-v\right)!}\right\} ,
\end{array}\label{eq:psdfs-ps}
\end{equation}
respectively. These analytical expressions are obtained using the
higher-order moment defined in Eqs. (6) and (7) of Ref. \cite{malpani2019lower}.
Similarly, the other parameter required in Eq. (\ref{eq:PE}) for
PADFS and PSDFS are obtained as 
\begin{equation}
\begin{array}{lcl}
\left(\frac{d\langle\hat{J}_{z}\rangle}{d\phi}\right)_{+} & = & \frac{1}{2}\frac{\left|N_{+}\right|^{2}}{n!}\sum\limits _{p,p'=0}^{n}{n \choose p}{n \choose p'}(-\alpha^{\star})^{(n-p)}(-\alpha)^{(n-p')}\exp\left[-\mid\alpha\mid^{2}\right] \sum\limits _{m=0}^{\infty}\frac{\alpha^{m}(\alpha^{\star})^{m+p-p'}(m+p+u)!}{m!(m+p-p')!}\left(m+p+u\right)\sin\text{\ensuremath{\phi}}
\end{array}\label{eq:pea}
\end{equation}
and
\begin{equation}
\begin{array}{lcl}
\left(\frac{d\langle\hat{J}_{z}\rangle}{d\phi}\right)_{-} & = & \frac{1}{2}\frac{\left|N_{-}\right|^{2}}{n!}\sum\limits _{p,p'=0}^{n}{n \choose p}{n \choose p'}(-\alpha^{\star})^{(n-p)}(-\alpha)^{(n-p')}\exp\left[-\mid\alpha\mid^{2}\right] \sum\limits _{m=0}^{\infty}\frac{\alpha^{m}(\alpha^{\star})^{m+p-p'}(m+p)!(m+p)!}{m!(m+p-p')!\left(m+p-v\right)!}\left(m+p-v\right)\sin\text{\ensuremath{\phi}},
\end{array}\label{eq:pes}
\end{equation}
respectively.

\end{document}